\documentclass[11pt]{article}

\usepackage{geometry}
\newcommand{\be}{\begin{equation}}
\newcommand{\ee}{\end{equation}}
\usepackage{amsmath}
\usepackage{amssymb}
\usepackage{latexsym}
\usepackage{epsfig}

\usepackage[vcentermath]{youngtab}

\usepackage{amsmath,amssymb,amsbsy}
\usepackage{latexsym,layout,amsfonts,amssymb,fontenc,xcolor,amsthm,multirow,amsfonts,bm,textcomp}
\usepackage{hyperref}
\usepackage{empheq}
\usepackage{accents}

\newsavebox{\uuunit}
\sbox{\uuunit}
    {\setlength{\unitlength}{0.825em}
     \begin{picture}(0.6,0.7)
        \thinlines
        \put(0,0){\line(1,0){0.5}}
        \put(0.15,0){\line(0,1){0.7}}
        \put(0.35,0){\line(0,1){0.8}}
       \multiput(0.3,0.8)(-0.04,-0.02){12}{\rule{0.5pt}{0.5pt}}
     \end {picture}}

\def\2{\frac12}
\def\4{\frac14}

\newcommand{\vev}[1]{\langle{#1}\rangle}

\def\equationautorefname~#1\null{eq.~(#1)\null
}

\hypersetup{
    bookmarks=false,         
    unicode=false,          
    pdftoolbar=true,        
    pdfmenubar=true,        
    pdffitwindow=false, 
    pdfstartview={FitH},    
    pdftitle={Dual graviton and (dual) Double field theory},    
    pdfauthor={Eric A.Bergshoeff, Olaf Hohm, Victor A. Penas, Fabio Riccioni},     
    pdfsubject={},   
    pdfnewwindow=true,      
    colorlinks=false,       
    linkcolor=blue,          
    citecolor=blue,        
    filecolor=blue,      
    urlcolor=blue           
    linkbordercolor={blue},
    citebordercolor={purple},
    urlbordercolor={gray}
}

\begin{document}

\begin{titlepage}
\begin{center}

\hfill UG-16-03 \\

\vskip 1.5cm

{\Large \bf
Dual Double Field Theory
}

\vskip 1.5cm

{\bf  Eric A.~Bergshoeff,$^1$ Olaf Hohm,$^2$ Victor A. Penas,$^1$ Fabio Riccioni$^3$}

\vskip 30pt

{\em $^1$ \hskip -.1truecm Centre for Theoretical Physics,
University of Groningen, \\ Nijenborgh 4, 9747 AG Groningen, The
Netherlands \vskip 5pt }

\vskip 15pt

{\em $^2$ \hskip -.1truecm Simons Center for Geometry and Physics, \\
Stony Brook University, \\
Stony Brook, NY 11794-3636, USA \vskip 5pt }

\vskip 15pt

{\em $^3$ \hskip -.1truecm
 INFN Sezione di Roma,   Dipartimento di Fisica, Universit\`a di Roma ``La Sapienza'',\\ Piazzale Aldo Moro 2, 00185 Roma, Italy
 \vskip 5pt }

\vskip 0.5cm

\small{e.a.bergshoeff@rug.nl, ohohm@scgp.stonybrook.edu, v.a.penas@rug.nl,\\
Fabio.Riccioni@roma1.infn.it}

\vskip 1cm

\end{center}

\vskip 0.5cm

\begin{center} {\bf ABSTRACT}\\[3ex]
\end{center}

We present the dual formulation of double field theory at the linearized level.
This is a classically equivalent theory describing the duals of the dilaton, the Kalb-Ramond field
and the graviton in a T-duality or $O(D,D)$ covariant way.
In agreement with previous proposals, the resulting theory encodes fields in mixed Young-tableau
representations, combining them into an antisymmetric 4-tensor under $O(D,D)$.
In contrast to previous proposals, the theory also requires an antisymmetric 2-tensor
and a singlet, which are not all pure gauge. The need for these additional fields is analogous
to a similar phenomenon for
``exotic" dualizations, and we clarify this by comparing with the dualizations of the
component fields. We close with some speculative remarks on the significance
of these observations for the full non-linear theory yet to be constructed.

\end{titlepage}

\newpage
\setcounter{page}{1} \tableofcontents


\setcounter{page}{1} \numberwithin{equation}{section}

\section{Introduction}

The duality transformations relating a field strength to its Hodge-dual, interchanging
Bianchi identities and field equations, are ubiquitous in gauge theory,
supergravity and string theory. For instance, the electromagnetic duality in four dimensions is essential
for the S-duality of ${\cal N}=4$ super-Yang-Mills theories. Moreover, in order to
define the world-volume dynamics of certain branes, it is necessary to replace some of the
standard $p$-form gauge potentials of string theory by their duals.

For a $p$-form potential, this dualization is straightforward: one simply replaces
its $(p+1)$-form field strength by the Hodge-dual of the field strength of the dual $(D-p-2)$-form.
For instance, the Kalb-Ramond 2-form $B_2$ of closed string theory in $D=10$
can be dualized to a 6-form $B_6$, which in turn couples to the NS5-brane.
Therefore, from the point of view of the full (non-perturbative) string or M-theory neither the 2- nor 6-form
is more fundamental, suggesting that a democratic formulation in which they
appear on equal footing is more appropriate.

Remarkably, taking into account further dualities or symmetries of string theory,
such as T-duality, then implies that even more fields of a more exotic nature
are needed. For instance, under the T-duality group $O(D,D)$ the 2-form $B_2$ transforms
into the metric. In fact, in double field theory, which makes the $O(D,D)$ symmetry manifest,
the metric and 2-form are part of an irreducible object, a generalized metric or
generalized frame. Thus, when dualizing $B_2$ into $B_6$, $O(D,D)$ covariance
requires that we also dualize the graviton into a `dual graviton'.
While at the linearized
level there is a straightforward procedure to dualize the graviton \cite{Hull:2000zn,West:2001as}, leading to a field
living in the mixed-Young tableau representation $(D-3,1)$,\footnote{We denote by $(p,q)$ the irreducible $GL(D)$
representation described by a Young diagram with two columns of lengths $p$ and $q$, where $p\geq  q$.}
there are strong no-go theorems
implying that at the non-linear level some new ingredients are needed \cite{Bekaert:2002uh,Bekaert:2004dz}.

In this paper we perform the dualization of double field theory (DFT) \cite{Siegel:1993th,Hull:2009mi,Hohm:2010pp} (see \cite{Aldazabal:2013sca} for reviews) at the linearized
level, thereby capturing in particular the dual potential $B_6$ and the dual graviton
in a T-duality covariant way. While we restrict ourselves to the free, quadratic theory,
we believe that our results give important pointers for the full non-linear theory.
The construction of the non-linear theory would be necessary in order to describe,
for instance, the world-volume dynamics of Kaluza-Klein monopoles (in a way
that is compatible with T-duality), for which the dual graviton is expected
to play the same role as the $B_6$ potential does for the NS5-brane \cite{Eyras:1998hn}.
More generally, one expects from T-duality the appearance of further mixed-Young
tableaux fields as exotic duals  \cite{Hull:2001iu,deMedeiros:2002qpr} of the usual gauge potentials \cite{Bergshoeff:2010xc},
as further clarified in \cite{Chatzistavrakidis:2013jqa}, together with associated `exotic' branes \cite{deBoer:2012ma}.
We find that all expected fields
are indeed described by the dual DFT.

The results of previous studies suggest that all dual fields can
be organized into a 4-index antisymmetric tensor under $O(D,D)$.  A first attempt to introduce a 4-index antisymmetric tensor into the DFT action (together with sources) was performed in \cite{Geissbuhler:2013uka} in the formulation of \cite{Geissbuhler:2013uka,Geissbuhler:2011mx}.
Moreover,
in \cite{Bergshoeff:2015cba} it was argued that the duality relation between the embedding tensor $\theta_{MNP}$ of lower dimensional supergravity and a $(D-1)$-potential $D_{(D-1),MNP}$ can be uplifted to higher dimensions by introducing mixed-symmetry dual potentials (in particular, relating the so-called $Q$- and $R$-fluxes to
$(8,2)$ and $(9,3)$ mixed-symmetry tensor fields, respectively) and that these mixed-symmetry potentials can be encoded in an antisymmetric 4-index tensor of $O(D,D)$.

Starting from the linearization of the DFT action, written in terms of the linearized
frame field that reads $h_{AB}=-h_{BA}$, we apply the standard procedure of obtaining the
dual theory, introducing Lagrange multiplier fields that impose the Bianchi identities for the generalized anholonomy coefficients (sometimes referred to as
generalized fluxes). This naturally leads to a 4-index antisymmetric field $D_{ABCD}$, but also to a field $D_{AB}$ in the antisymmetric 2-tensor representation and a field $D$ in
the singlet representation, in the following called $D$-fields. In this paper
the $D$-fields carry flat indices
$A,B=1,\ldots,2D$ under
the doubled local Lorentz group $O(D-1,1)\times O(D-1,1)$. The fields $D_{AB}$ and $D$
carry the same representations and gauge transformations as the
(linearized) generalized frame and the dilaton of the original DFT.
They are \textit{not} pure gauge under the (doubled) local
Lorentz symmetry (unlike, say, the antisymmetric part of the linearized vielbein in Einstein gravity),
and hence it seems inevitable to introduce a `second copy' of the original DFT fields in order to
formulate a duality- and gauge-invariant theory for the dual fields.

This observation is
the main unexpected result of our investigation, but it turns out that, upon reducing
to the physical, `undoubled' spacetime and breaking $O(D,D)$ to $GL(D)$,
the dual theory and its fields can be matched precisely with what one should expect for
the dualization of the `component' fields (i.e.~without employing the DFT formalism).
This match requires  a careful analysis of so-called `exotic' dualizations \cite{Hull:2001iu,deMedeiros:2002qpr}, in which,
for instance, the Kalb-Ramond 2-form $B_2$ is not dualized into
a 6-form in $D=10$ but into a gauge field with $(8,2)$ Young tableau symmetry.
The precise dynamical implementation
of such dualizations has only been investigated quite recently, in the work
of Boulanger et~al.~\cite{Boulanger:2012df}.
One of the novel features of such dualizations is that an off-shell formulation (i.e.~an action) only exists
provided extra fields are included which, however, are non-propagating and nicely
fit into the spectrum of representations determined before by independent methods.

Given the necessity of extra fields for exotic dualizations, it is perhaps not surprising that
we encounter extra fields in the dualization of DFT (which, again, do not upset the counting
of degrees of freedom), but similar features have also been encountered
in the extension of DFT to U-duality groups.
In this so-called  `exceptional field theory' (EFT) the inclusion of (parts of) the dual graviton
is unavoidable \cite{Hohm:2013jma}. In EFT, the extra
fields associated to the dual graviton satisfy
unusual constraints, but they do allow for a formulation including parts of the
dual graviton at the full non-linear level.\footnote{Moreover, these extra fields are
necessary for supersymmetry \cite{Godazgar:2014nqa} and in order for generalized Scherk-Schwarz
compactifications to be consistent \cite{Hohm:2014qga}.} We hope to return to the problem of understanding
the precise relation of the extra fields found in DFT and EFT in the presence of the dual
graviton and at the full non-linear level.\footnote{In the E$_{11}$ proposal of \cite{West:2001as} the dual fields $D_{ABCD}$ emerge
naturally under a level-decomposition w.r.t.~the $O(10,10)$ subgroup. The additional
fields seem to be absent. We refrain from speculating about the significance
of this observation for the E$_{11}$ program.}

The rest of this paper is organized as follows.
To set the stage for the dualization of DFT, in sec.~2 we review the
standard dualization of $p$-form gauge potentials and the graviton
at the linearized level. Moreover, we work out the dualization in `string frame',
i.e., in gravity plus dilaton, which shows some important differences to the
dualization in Einstein frame.  In sec.~3 we turn to `exotic' dualizations,
and we discuss in detail the dualization of the Kalb-Ramond 2-form to a  $(D-2,2)$ potential
plus extra fields.  In sec.~4 we perform the dualization for linearized DFT.
The geometric content of the dual DFT action is discussed in sec.~5.
In sec.~6 we compare the DFT results with the component
results and find precise agreement.
We close with some general remarks and speculations on the non-linear theory in
the conclusion section.

\section{Standard Dualizations}

\subsection{$p$-form dualization}
As a warm-up we start by recalling  the dualization of the electromagnetic field in four dimensions. Starting with the Maxwell action
\begin{equation}
S[A] \ = \ -\frac{1}{4}\int d^4 x \,F_{ab}\, F^{ab} \; ,
\end{equation}
where $F_{ab} = 2\partial_{[a} A_{b]}$, one moves to a first-order formulation where $F_{ab}$ is an independent field, and the Bianchi identity is imposed by introducing a Lagrange multiplier $\tilde{A}_a$,
\begin{equation}
S[A,F] \ = \ \int d^4 x \Big( - \frac{1}{4} F_{ab}F^{ab} + \frac{1}{2}\epsilon^{abcd} \tilde{A}_a \partial_b F_{cd} \Big) \; . \label{maxwellduality}
\end{equation}
This action is gauge invariant under $\delta \tilde{A}_a = \partial_a \Lambda$, $\delta F_{ab}=0$.
Varying w.r.t.~$\tilde{A}_a$ one obtains the Bianchi identity $\partial_{[a}F_{bc]}=0$, which can be solved
in terms of the Maxwell potential, giving back  the original Maxwell theory.
Conversely, one can solve for $F$ in terms of $\tilde{A}_a$ to obtain the duality relation
\begin{equation}
F_{ab} = \frac{1}{2} \epsilon_{ab}{}^{cd} \tilde{F}_{cd} \; ,
\end{equation}
where $\tilde{F}_{ab} = 2\partial_{[a} \tilde{A}_{b]}$ is the dual field strength. Insertion into the action leads to the dual Maxwell action for $\tilde{F}_{ab}$.

By applying the same procedure in any dimension and for any $p$-form potential $A_p$, one obtains a dual $(D-p-2)$-form potential $\tilde{A}_{D-p-2}$, whose gauge paramenter is a $(D-p-3)$-form,
\begin{equation}
\delta \tilde{A}_{D-p-2} = d \Lambda_{D-p-3} \; .
\end{equation}
In order to set the stage for the comparison with the dualization in DFT, we will often consider
the Hodge duals of the potential $\tilde{A}_{D-p-2}$ and the gauge parameter $\Lambda_{D-p-3}$.
The corresponding field is denoted by $\tilde{A}_{p+2}$ and the parameter by $\Lambda_{p+3}$.
The field strengths and the gauge variation then take the divergence form
\begin{equation}
F^{a_1\ldots a_{p+1}} \ = \ \partial_b\tilde{A}^{ba_1\ldots a_{p+1}}\;, \qquad
\delta \tilde{A}^{a_1 ...a_{p+2}} = \partial_{a} \Lambda^{a a_1 ...a_{p+2}}\;,
\end{equation}
while the corresponding first-order action reads
  \begin{equation}
 S[\tilde{A}_{p+2},F_{p+1}] \ = \ \frac{1}{(p+1)!} 
  \int d^D x \Big( - \frac{1}{2 } F_{a_1 ... a_{p+1}}F^{a_1 ... a_{p+1}} - \tilde{A}^{a_1 ... a_{p+2}} \partial_{a_1} F_{a_2 ...a_{p+2}} \Big) \; .
\end{equation}

For instance, consider a 2-form $b_2$ in $D$ dimensions with field strength $H_{abc}=3\partial_{[a}b_{bc]}$.
Starting from the standard action
\begin{equation}
S[b] \ = \ -\frac{1}{12}\int d^Dx \,H_{abc}\,H^{abc}\,,
\end{equation}
we pass to a first order action with a fully antisymmetric $4$-tensor $D^{abcd}$ and 3-form $H_{abc}$ as independent fields,
\begin{equation}\label{masteractionHtheory}
S[D,H] \ = \
\int d^Dx\Big(-\frac{1}{12}H_{abc}H^{abc}+ D^{abcd}\partial_{a}H_{bcd}\Big) \;.
\end{equation}
The equation for $D^{abcd}$ gives the Bianchi identity for $H$, $\partial_{[a}H_{bcd]}=0$, while the equation for $H$ gives the duality relation
\begin{equation}\label{dualityKR}
-\tfrac{1}{6} H^{abc} \ = \ \partial_{d}D^{dabc} \;.
\end{equation}
The action and field equations are invariant under the gauge transformation
\begin{equation}\label{SigmaforDabcdHtheory}
\delta D^{abcd}=\partial_e\Sigma^{eabcd}\,,
\end{equation}
where $\Sigma^{eabcd}$ is completely antisymmetric.
The more familiar form of the duality relation is obtained by passing to the Hodge-dual $(D-4)$-form
 \be
  \tilde{D}_{a_1\ldots a_{D-4}} \ \equiv \ \tfrac{1}{4!}\,\epsilon_{a_1\ldots a_{D-4}b_1\ldots b_4}\, D^{b_1\ldots b_4}\;,
 \ee
in terms of which (\ref{dualityKR}) reduces to the standard duality relation between the $(D-3)$-form field strength of this $(D-4)$-form potential
and $H$.  Alternatively,
defining
\begin{equation}\label{tildeHintermsofGabc}
\tilde{H}_{abc}\ \equiv \ -2\eta_{ad}\eta_{be}\eta_{cf}G^{def} \ \equiv \
-6\eta_{ad}\eta_{be}\eta_{cf}\partial_gD^{gdef},
\end{equation}
the above duality relation
reads $H_{abc}=\tilde{H}_{abc}$. The  `field strength' $G^{abc}$ in the above equation will appear naturally
in Section 6.
The equations of motion and Bianchi identity for the dual field are then swapped with respect to the original variables:
\begin{align}\label{eomtildeH}
&(\mbox{E.o.M})~~\partial_{[a}\tilde{H}_{bcd]} \ = \ 0\;,\\[8pt]
&(\mbox{B.I})~~~~~~~\partial_a\tilde{H}^{abc} \ = \ 0 \; .\label{BIoftildeH}
\end{align}
%


\subsection{The dual graviton}

We now repeat the same analysis for the dual of the $D$-dimensional graviton at the 
linearised level, following \cite{West:2001as,Boulanger:2003vs}. 
We write the linearized Einstein-Hilbert action for the vielbein 
fluctuation $h_{a|b}$ (including the antisymmetric part,
as indicated by the bar) as
\begin{equation}\label{linEH}
S_{\rm EH}[h] \ = \ \int d^D x \,\big[ f_{ab}{}^b f^{ac}{}_c - \tfrac{1}{2} f_{abc} f^{acb} - \tfrac{1}{4} f_{abc} f^{abc}\,\big ]\;,
\end{equation}
with the linearized coefficients of anholonomy,
\begin{equation}\label{ffluxdefinition}
 f_{ab}{}^c \ = \ 2 \partial_{[a}h_{b]|}{}^c\,.
 \end{equation}
These quantities satisfy the Bianchi identity
\begin{equation}
\partial_{[a} f_{bc]}{}^d \ = \ 0 \; ,\label{Bianchigraviton}
\end{equation}
while the field equations obtained by variation w.r.t~$h$ are
\begin{equation}\label{linEinstein}
\partial^c f_c{}_{(ab)} +\partial_{(a}f_{b)c}{}^{c} -\eta_{ab}\,\partial^c f_{cd}{}^{d}\ = \ 0 \; .
\end{equation}

We now pass to a first order action by adding the Lagrange
multiplier $D^{abc}{}_d\equiv D^{[abc]}{}_d$ to impose the Bianchi identity,
\begin{align}\label{masteractiongrav}
S[f,D] \ = \
\int d^Dx\,\Big(  f_{ab}{}^b f^{ac}{}_c - \tfrac{1}{2} f_{abc} f^{acb} - \tfrac{1}{4} f_{abc} f^{abc}
+3D^{abc}{}_d\, \partial_{a}f_{bc}{}^d \Big)\;.
\end{align}
Varying w.r.t.~$D^{abc}{}_d$ and $f_{ab}{}^c$, respectively, gives
\begin{align}\label{dualGravRel}
&\partial_{[a}f_{bc]}{}^d \ = \ 0,\\[5pt]
&-\tfrac{1}{2}f^{ab}{}_c-f^{[a}{}_c{}^{\,b]}- 2 \delta_c{}^{[a} f^{b]}{}_d{}^d \ = \
3\, \partial_dD^{dab}{}_c \; . \nonumber
\end{align}
The first equation implies locally that $f$ takes the form (\ref{ffluxdefinition}).
The second equation is then the duality relation between the graviton, contained in $h_{a|}{}^{b}$,
and the dual graviton, contained in $D^{abc}{}_{d}$. From this duality relation we may recover the
original (linearized) Einstein equations (\ref{linEinstein})
by acting with $\partial_a$ and
using that the right-hand side gives zero by the `Bianchi identity'
$\partial_d\partial_aD^{dab}{}_{c}\equiv 0$.

Conversely, we can express the theory in terms of the dual variables.
We first note that in terms of the `field-strength' for the dual graviton,
\begin{equation}
 G_{a}{}^{bc} \ \equiv \ 3\,\partial_{d}D^{dbc}{}_a\ \,,
\end{equation}
the duality relation is equivalent to
\begin{equation}\label{gabcintermsofGabconlygravity}
f_{ab}{}^c \ =  \  2\,G_{[ab]}{}^c -\frac{2}{D-2}  G_d{}^d{}_{[a} \delta_{b]}{}^{c}
 \ = \ 6\,\partial_eD^{e}{}_{[b}{}^{c}{}_{a]} -\frac{6}{D-2} \, \partial_e D^e{}_{d[a}{}^d \,\delta_{b]}{}^c\,,
\end{equation}
where we  reinserted the explicit potentials in the last step.
Inserting now this expression for $f$ in terms of $D$ into (\ref{masteractiongrav}) one
obtains the dual action for $D$.

Let us discuss the physical content of the dual theory in a little more detail.
To this end we decompose
\begin{equation}\label{splitDhatgravandtrace}
D^{abc}{}_d \ = \ D^{(tr)abc}{}_d \ + \ 3\,\delta_d{}^{[a}D'^{bc]} \ ,
\end{equation}
where $D^{(tr)abc}{}_d$ is traceless and $D'^{ab}=\frac{1}{(D-2)}D^{abc}{}_c$ is the trace part.
In order to further elucidate the representation content, consider the `Hodge-dual' field
 \be
  \tilde{D}_{a_1\ldots a_{D-3}|b} \ \equiv \ \tfrac{1}{6}\,\epsilon_{a_1\ldots a_{D-3}cde}\, D^{cde}{}_{b}\;,
 \ee
whose irreducible $GL(D)$ representations are given by
 \be
  (D-3) \; \otimes \;   {\tiny \yng(1)} \; = \; (D-3,1) \; \oplus \; (D-2)\;.
 \ee
It is easy to see that the traceless potential $D^{(tr)abc}{}_d$ in eq.~(\ref{splitDhatgravandtrace}) corresponds to
the $(D-3,1)$ mixed Young-tableaux representation, while  $D'^{ab}$ corresponds to
the totally antisymmetric $(D-2)$.  It turns out that the totally antisymmetric representation is pure gauge. Indeed,
the gauge invariance of the linearized Einstein-Hilbert action (\ref{linEH}) under
diffeomorphisms and local Lorentz transformations
\begin{equation}
\delta h_{a|b}=\partial_a\xi_b-\Lambda_{ab}\,,
\end{equation}
elevates to a gauge invariance of the master action (\ref{masteractiongrav}), acting
on the fields (\ref{splitDhatgravandtrace}) as
  \begin{equation}\label{localLorentz}
 \delta_{\Lambda}D^{(tr)abc}{}_d \ = \ 0\;, \qquad
 \delta_\Lambda D'^{ab} \ = \  \tfrac{1}{3}\,\Lambda^{ab} \ .
\end{equation}
Due to this St\"uckelberg invariance, the field $D'^{ab}$ drops out of the action upon insertion of (\ref{gabcintermsofGabconlygravity}) into (\ref{masteractiongrav}), leaving a two-derivative action for
the physical dual graviton in the $(D-3,1)$ Young tableau representation.\footnote{It should be emphasized
that while the standard Einstein-Hilbert action can be written entirely in terms of the symmetric $h_{(ab)}$
and the dual action entirely in terms of the irreducible $(D-3,1)$, the dualization requires the presence
of an antisymmetric part, either $h_{[ab]}$ or $D'^{ab}$, since in the master action (\ref{masteractiongrav})
or the duality relations (\ref{dualGravRel}), local Lorentz invariance  allows us to set
only one to zero, not both.}
The $D$-fields also possess gauge transformations that leave the `field strength' $G_{a}{}^{bc}$ and hence
the action and duality relations invariant,
\begin{equation}
 \delta_\Sigma D^{abc}{}_d \ = \ \partial_e\Sigma^{eabc}{}_d \;,
\end{equation}
with the parameter $\Sigma^{eabc}{}_d=\Sigma^{[eabc]}{}_d$ (that could be decomposed
into traceless and trace part in order to obtain the gauge transformations of $D^{(tr)abc}{}_d$ and $D'^{ab}$).


\subsection{Dual graviton and dilaton}

We now consider the dual graviton and  dilaton  together at the linearized level. We first consider
a canonically coupled scalar (i.e.~in Einstein frame) with Lagrangian
${\cal L}=R-\tfrac{1}{2}(\partial \varphi)^2$. We thus add to the linearized action (\ref{masteractiongrav}) the first-order
action\footnote{The superscript E refers to the Einstein-frame.}
\begin{align}\label{firstscalar}
S[f^{(E)}_a,D^{(E) ab}] \ = \ \int d^Dx\Big( -\frac{1}{2} f^{(E)a} f^{(E)}_a  + D^{(E) ab} \partial^{}_a f^{(E)}_b \Big) \ ,
\end{align}
where the antisymmetric $D^{(E) ab}$ is the Lagrange multiplier whose equation of motion
yields the Bianchi identity
\begin{equation}
\partial^{}_{[a} f^{(E)}_{b]} \ = \ 0\;.
\end{equation}
This implies locally $f^{(E)}_a = \partial_a \varphi $, from which we recover upon reinsertion
into (\ref{firstscalar}) the original scalar theory.
Alternatively, varying w.r.t.~$f^{(E)}_a$ gives the duality relation
\begin{equation}
f^{(E)a} \ = \ \partial_b D^{(E) ab}
\; ,
\end{equation}
and eliminating $f^{(E)}_a$ accordingly from (\ref{firstscalar})
yields the theory for the dual dilaton $D^{(E) ab}$ (or, equivalently, for the $(D-2)$-form potential).
The action and duality relations are invariant under the gauge tranformation
\begin{equation}
\delta D^{(E) a b} = \partial_c \Sigma^{cab}\;,
\end{equation}
where $\Sigma^{abc}$ is fully antisymmetric.

The above dualization of a scalar was completely decoupled from the dualization of gravity.
For dualization in string frame, however, this picture changes significantly in that it will be the trace of the field
$D^{abc}{}_d$ (c.f.~the previous subsection) that becomes the dual dilaton, while the analogue of
$D^{(E) ab}$ will be pure gauge, transforming with a shift under local Lorentz transformations.

We start from the action with Lagrangian ${\cal L}=e^{-2\phi}(R+4(\partial\phi)^2)$ for the graviton-dilaton sytem,
whose linearization yields
  \begin{equation}\label{actiongravanddil}
S[h_{a|b},\phi] \ = \ \int d^Dx\left(f^af_a-\frac{1}{4}f_{ab}{}^cf^{ab}{}_c-\frac{1}{2}f_{ab}{}^cf^{a}{}_c{}^b\right),
\end{equation}
where
\begin{equation}\label{calF}
f_a \ \equiv \ f_{ab}{}^b+2\,\partial_a\phi\;,
\end{equation}
with the coefficients of anholonomy defined in (\ref{ffluxdefinition}).
Varying w.r.t.~the vielbein and the dilaton, respectively, one obtains the
equations of motion
\begin{equation}\label{eomdilgravanddil}
\partial_cf^c{}_{(ab)}+\partial_{(a}f_{b)} \ = \ 0\;,\qquad
\partial_af^a \ = \ 0\;.
\end{equation}
The $f_{ab}{}^{c}$ and $f_a$ satisfy the following Bianchi identities:
\begin{equation}\label{contractedBIgravanddil}
\partial_{c}f_{ab}{}^c+2\,\partial_{[a}f_{b]} \ = \ 0\; ,\qquad
\partial_{[a}f_{bc]}{}^d \ = \ 0\;.
\end{equation}
As before, we can pass to a first-order action with Lagrange multipliers $D^{abc}{}_d=D^{[abc]}{}_d$ and $D'^{ab}=D'^{[ab]}$ imposing the Bianchi identities,
\begin{align}\label{masteractiongravanddil}
S[f_a, f_{ab}{}^c, D,D^{\prime}] \ = \
\int d^Dx\Big( \, &f^af_a-\frac{1}{4}f_{ab}{}^cf^{ab}{}_c-\frac{1}{2}f_{ab}{}^cf^{a}{}_c{}^b
\\ \nonumber
&\;+3D^{abc}{}_d\, \partial_{a}f_{bc}{}^d+
D'^{ab}\big(\partial_cf_{ab}{}^c+2\partial_{ a}f_{b }\big)\Big).
\end{align}
Varying w.r.t~the fundamental fields $D^{abc}{}_d$,  $D'^{ab}$, $f_a$ and $f_{ab}{}^c$, respectively,
one obtains
\begin{align}\label{delDhatgravanddil}
&\partial_{[a}f_{bc]}{}^d \ = \ 0,\\[5pt] \label{delDprimegravanddil}
&\partial_c f_{ab}{}^c+2\partial_{[a}f_{b]} \ = \ 0, \\[5pt] \label{delFagravanddil}
&f^a \ = \ \partial_{b}D'^{ba},\\ \label{delFabcgravanddil}
&-\frac{1}{2}f^{ab}{}_c-\frac{1}{2}f^{a}{}_c{}^b+\frac{1}{2}f^b{}_c{}^a \ = \ 3\partial_eD^{eab}{}_c+\partial_{c}D'^{ab}\;.
\end{align}
It is straightforward to see, using the Poincar\'e lemma, that the general solution of the first two equations, (\ref{delDhatgravanddil}) and (\ref{delDprimegravanddil}), give back (\ref{ffluxdefinition}) and (\ref{calF}),
which upon back-substitution into the action gives the string frame action (\ref{actiongravanddil})
for dilaton plus gravity.
The final two equations above, (\ref{delFagravanddil}) and (\ref{delFabcgravanddil}), are duality relations, which
allow us to recover the second-order equations of motion (\ref{eomdilgravanddil}) as integrability conditions.
To this end we act on (\ref{delFagravanddil}) with $\partial_{a}$, which by the Bianchi identity $\partial_a\partial_b D^{\prime ab}\equiv 0$ implies
\begin{equation}\label{scalargravanddileom}
\partial_af^a \ = \ 0~~\Leftrightarrow~~\partial_{a}f^a{}_b{}^b+2\,\partial^a\partial_a\phi \ = \ 0\;,
\end{equation}
in agreement with the dilaton field equation (the second equation in (\ref{eomdilgravanddil})). In order to obtain the first equation in (\ref{eomdilgravanddil}) we act with $\partial_a$ on (\ref{delFabcgravanddil}) to obtain
\begin{equation}\label{hitwithpartialdduality2}
\begin{split}
-\frac{1}{2}\partial_af^{ab}{}_c-\frac{1}{2}\partial_af^a{}_c{}^b+\frac{1}{2}\partial_af^b{}_c{}^a
\ = \ \partial_c\partial_aD'^{ab} \ = \ \partial_cf^b\;,
\end{split}
\end{equation}
using in the last step the first duality relation (\ref{delFagravanddil}). After lowering the index $b$ and symmetrizing in $(b,c)$, equation (\ref{hitwithpartialdduality2}) becomes equivalent to the first equation in (\ref{eomdilgravanddil}).
Note that the antisymmetric combination
in $(b,c)$ is zero by the Bianchi identity (\ref{contractedBIgravanddil}).
Thus, we have correctly recovered the equation of motion for the graviton.

We can also solve eqs.~(\ref{delFagravanddil}) and (\ref{delFabcgravanddil}) for $f_a$ and $f_{ab}{}^c$ in terms
of the $D$-fields. Back-substitution into (\ref{masteractiongravanddil}) then yields the dual theory,
which we analyze now in a little more detail.
Defining the dual field strengths
\begin{equation}\label{fieldstrengthgravanddil}
 G_{a}{}^{bc} \ \equiv \ 3\, \partial_{e}D^{ebc}{}_a+\partial_{a}D'^{bc}\;, \qquad
 g^{a} \ \equiv \ \partial_bD'^{ba}\;,
\end{equation}
we find
\begin{equation}\label{dualityrelationintermsofgabc}
 f_{ab}{}^c \ = \ g_{ab}{}^{c} \ \equiv \ 2G_{[ab]}{}^c\;, \qquad
f_a \ = \ g_a\;,
\end{equation}
where we introduced $g_{ab}{}^{c}$ and $g_a$ for convenience.
The equations of motion and Bianchi identities for the dual system are then
\begin{equation}\label{equationsgabcandGagravanddil}
\mbox{B.I's:}\left\{
  \begin{array}{lr}
   \partial_{c}g^c{}_{(ab)} + \partial_{(a}g_{b)}=0\,, \\[10pt]
   \partial_cg_{ab}{}^c+2\partial_{[a}g_{b]}=0\,,\\ [10pt]
	   \partial_ag^a=0\,.
		 \end{array}
\right.
~~~~~~~~~
\mbox{E.o.M's:}~\left\{~\partial_{[a}g_{bc]}{}^d=0\,.
\right.
\end{equation}

In order to further analyze the content of these equations it
is useful to decompose $D$ as follows
\begin{equation}\label{splitDhatgravanddil}
D^{abc}{}_d \ = \ D^{(tr)abc}{}_d+3\,\delta_d{}^{[a}\bar{D}^{bc]}\;,
\end{equation}
where $D^{(tr)abc}{}_d$ is traceless and $\bar{D}^{ab}=\bar{D}^{[ab]}$ the trace part.
The equations of motion for the components then read
\begin{align}
 \partial_e\partial_{[a}D^{(tr)e}{}_{b\,}{}^d{}_{c]} \ = \ 0\;, \qquad
 \partial_c\partial_{[a}\bar{D}^c{}_{b]} \ = \ 0 \;.
\end{align}
Note that the $D'^{ab}$'s dropped out, which means that they are subject to a St\"uckelberg symmetry.
From the duality relations (\ref{dualityrelationintermsofgabc}) and the split (\ref{splitDhatgravanddil}) it is easy to obtain the usual duality relation between the dilaton and the dual dilaton:
\begin{equation}
f_a-f_{ab}{}^b=g_a-g_{ab}{}^b~\Rightarrow~ 2\partial_a\phi= 3(D-2)\partial_c\bar{D}^c{}_a.
\end{equation}
We observe that $D'^{ab}$ disappears and the field $\bar{D}^{ab}$ (the trace of $D^{abc}{}_d$) is the dual dilaton,
which is the opposite of the situation in Einstein frame.

We close this subsection by discussing the gauge transformations for the $D$-fields. The duality relations (\ref{delFagravanddil}) and the master action are invariant under local Lorentz transformations with
$D'^{ab}$ and  $D^{abc}{}_d$ transforming as
\begin{equation}
 \delta_\Lambda D^{abc}{}_d \ =  \ 0\;, \qquad
 \delta_\Lambda D'^{ab} \ = \ \Lambda^{ab}\;.
\end{equation}
W.r.t~to the decomposition (\ref{splitDhatgravanddil}) this implies in particular
$\delta_\Lambda D^{(tr)abc}{}_d=0$ and $\delta_\Lambda \bar{D}_{ab}=0$, implying that the physical dual graviton and the dual dilaton are invariant. The $D$-fields also possess gauge transformations that leave the `field strengths'
$G_{a}{}^{bc}$ and $g^a$ (and hence the duality relations and action) invariant,
\begin{equation}
\delta_\Sigma D^{abc}{}_d \ = \ \partial_e\Sigma^{eabc}{}_d + \partial_d\Sigma^{abc}\;, \qquad
\delta_\Sigma D'^{ab} \ = \ -3\partial_e\Sigma^{eab}\;.
\end{equation}
The gauge parameters satisfy $\Sigma^{eabc}{}_d=\Sigma^{[eabc]}{}_d$ and $\Sigma^{abc}=\Sigma^{[abc]}$. One may decompose into traceless and trace parts in order to read off the transformations for $D^{(tr)abc}{}_{d}$ and $\bar{D}^{ab}$.



\section{Exotic Dualization of Kalb-Ramond field}

In this section we will discuss the dualization of a 2-form gauge potential
(`the $B$-field') into exotic mixed Young tableau fields. We first review general aspects of such mixed
Young tableau gauge fields and then turn to a master action that can be used to
dualize the $B$-field into such a tensor, provided extra fields are included.
These fields are quite unusual in that they are not auxiliary (they cannot be
eliminated algebraically) nor pure gauge, yet they do not add to the propagating
degrees of freedom.

\subsection{Generalities of $(D-2,2)$ Young tableaux gauge fields}
We start by discussing general aspects of gauge fields in mixed Young diagram representations; 
see \cite{Bekaert:2002dt,deMedeiros:2002qpr} for a systematic treatment and \cite{Siegel:1986tn} for the 
construction of invariant actions. Here we specialize to the $(D-2,2)$ Young diagram representation, for which 
the gauge field is subject to 
 \be
  B_{a_1\ldots a_{D-2},bc} \ \equiv \  B_{[a_1\ldots a_{D-2}],bc} \ \equiv \ B_{a_1\ldots a_{D-2},[bc]}\;, \qquad
  B_{[a_1\ldots a_{D-2},b]c} \ \equiv \ 0\;.
 \ee
There are two types of gauge parameters, $\mu\in (D-3,2)$ and $\lambda\in (D-2,1)$, 
acting as\footnote{We sometimes underline indices in order to indicate which indices participate in an 
antisymmetrization.} 
 \be\label{D-22Gauge}
 \begin{split}
  \delta B_{a_1\ldots a_{D-2},bc} \ = \ &(D-2)\,\partial_{[a_1}\mu_{a_2\ldots a_{D-2}],bc}\\[0.5ex]
  &+ \partial_{[\underline{b}}\lambda_{a_1\ldots a_{D-2},\underline{c}]}
  +\tfrac{1}{2}(D-2)\,\partial_{[a_1}\lambda_{|bc|a_2\ldots a_{D-3}, a_{D-2}]}\;.
 \end{split}
 \ee
These gauge transformations preserve the algebraic constraints on $B$.
We can define a gauge invariant curvature, starting from the first-order generalized
Christoffel symbol
 \be\label{geneChris}
  \Gamma_{a_1\ldots a_{D-1},bc} \ \equiv \ (D-1)\,\partial_{[a_1}B_{a_2\ldots a_{D-1}],bc}\;,
 \ee
which is invariant under $\mu$ transformations and satisfies the Bianchi identities
 \be\label{GammaBianchi}
  \Gamma_{[a_1\ldots a_{D-1},b]c} \ = \ 0\;, \qquad \partial_{[a_1}\Gamma_{a_2\ldots a_{D}],bc} \ = \ 0\;.
 \ee
As common for Young tableau fields
with more than one column, this first-order object is not fully gauge invariant (it is analogous to the
Christoffel symbols), because under $\lambda$ transformations
we have
 \be
  \delta_{\lambda}\Gamma_{a_1\ldots a_{D-1},bc} \ = \ (D-1)\, \partial_{[\underline{b}}
  \partial_{[a_1}\lambda_{a_2\ldots a_{D-1}],\underline{c}]}\;.
 \ee
A fully gauge invariant curvature is the Riemann-like tensor obtained by taking another
derivative and antisymmetrizing over three indices,
 \be
  {\cal R}_{a_1\ldots a_{D-1}, bcd} \ \equiv \
  3\,\partial_{[\underline{b}} \Gamma_{a_1\ldots a_{D-1},\underline{c}\underline{d}]}\;.
 \ee
This Riemann tensor satisfies the Bianchi identities
 \be
  {\cal R}_{ [a_1\ldots a_{D-1}, b]cd} \ = \ 0\;, \qquad
  \partial_{[a_1}{\cal R}_{a_2\ldots a_{D}],bcd} \ = \ 0\;,
 \ee
and hence lives in the $(D-1,3)$ Young diagram representation.

Naively, one would now impose
the Einstein-type field equations that set to zero the generalized Ricci tensor
${\cal R}_{a_1\ldots a_{D-2}}{}^{d}{}_{,bcd} $, but it turns out
that a theory with these field equations is actually topological.
To see this note that these field equations
imply vanishing of the double-trace of the Riemann tensor,
which by the equivalence
 \be
  {\cal R}_{a_1\ldots a_{D-3}}{}^{bc}{}_{,a_{D-2}bc} \ = \ 0 \quad \Leftrightarrow \quad
  \epsilon_{a_1\ldots a_{D-3}}{}^{cde}\,\epsilon_{a_{D-2}}{}^{b_1\ldots b_{D-1}}\,
  {\cal R}_{b_1\ldots b_{D-1},cde} \ = \ 0\;,
 \ee
implies vanishing of the full Riemann tensor and hence that the field is pure gauge.
However, we can impose weaker field equations that do lead to propagating degrees of freedom,
setting to zero the triple-trace of the Riemann tensor,
 \be\label{SCALAR}
  {\cal R}_{a_1\ldots a_{D-4}}{}^{bcd}{}_{,bcd} \ = \ 0\;.
 \ee
Note that these are the same number of equations as for the conventional dual of
a 2-form ($D-p-2=D-4$), but now these are equations for the $(D-2,2)$ gauge field.
(Such dualities have been discussed by Hull in \cite{Hull:2001iu}; see also \cite{Deser:1980fy} for 
similar exotic dualizations.) 
This also proves that there can be no action principle implying (\ref{SCALAR}) for the $(D-2,2)$ gauge field alone --- simply because variation w.r.t.~the $(D-2,2)$ field would yield more
equations. However, one can write an action that implies this field equation at the cost
of introducing more fields (that are not pure gauge), which also serves as a master action
proving the equivalence with the standard 2-form action, as we will now discuss.

\subsection{Master action}

In order to construct this master action we follow \cite{Boulanger:2012df} and
write the standard action for the Kalb-Ramond field up to total derivatives as
 \be\label{secondorderb}
  S[ b ] \ = \ -\tfrac{1}{12}\int d^Dx\,H^{abc} H_{abc} \ = \
  -\tfrac{1}{4}\int d^Dx\, \big(\partial^a b^{bc}\,\partial_a b_{bc}
  -2\,\partial_ab^{ab}\,\partial^c b_{cb}\big)\;.
 \ee
We can then replace it by the first-order action
 \be\label{firstorderY}
  S[Q,D] \ = \ \int d^Dx\big(-\tfrac{1}{4}\,Q^{a|bc}Q_{a|bc}+\tfrac{1}{2}\,Q_{a|}{}^{ab} Q^{c|}{}_{cb}
  -\tfrac{1}{2}\,D^{ab|cd}\,\partial_{a}Q_{b|cd}\big)\;,
 \ee
where the fields have the symmetries
 \be
  Q_{a|bc} \ = \ -Q_{a|cb}\;, \qquad D_{ab|cd} \ = \ -D_{ba|cd}  \ = \ -D_{ab|dc}\; .
 \ee
Note that, as usual for master actions, these fields do not live in irreducible representations.
The above action may seem like a rather unnatural rewriting of a 2-form theory,
but we will see in sec.~6 that, in the appropriate sector, DFT reproduces precisely such an action.
This action is invariant under the gauge transformations
 \be\label{YPgauge}
  \begin{split}
   \delta Q_{a|bc} \ &= \ \partial_a K_{bc}\;, \qquad K_{ab} \ \equiv \ 2\partial_{[a}\tilde{\xi}_{b]}\;, \\[0.5ex]
   \delta D_{ab,cd} \ &= \ \partial^e\Sigma_{eab|cd} +4\,\eta_{[a[\underline{c}}K_{b]\underline{d}]} \;,
  \end{split}
 \ee
where $\Sigma_{abc|de}\equiv \Sigma_{[abc]|de}\equiv \Sigma_{abc|[de]}$.

Let us now verify the equivalence with the second-order action. We vary w.r.t.~$D$ and $Q$, respectively,
to obtain
 \be\label{YPEqs}
 \begin{split}
  \partial_{[a}Q_{b]|cd} \ &= \ 0 \quad \Rightarrow \quad Q_{a|bc} \ = \ \partial_ab_{bc}\;, \\[0.5ex]
  \partial^d D_{da|bc} \ &= \ Q_{a|bc} - \eta_{ab}Q^{d}{}_{|dc} +  \eta_{ac}Q^{d}{}_{|db}\;.
 \end{split}
 \ee
Reinserting the solution of the first equation into the action we recover the original
second-order action (\ref{secondorderb}). Equivalently, at the level of the equations of motion,
we can  act on the second equation with $\partial^a$ to obtain the Bianchi identity
 \be\label{BianchiBeom}
  0 \ = \ \partial^a\partial^d D_{da|bc} \ = \ \partial^aQ_{a|bc} - \partial_bQ^{a}{}_{|ac}
  +  \partial_cQ^{a}{}_{|ab}
  \ = \ \partial^a\big(\partial_a b_{bc}-\partial_b b_{ac}+\partial_cb_{ab}\big)\;,
 \ee
which becomes the standard second-order equation for $b_{ab}$.
Thus, the first-order action is on-shell equivalent to the second-order action.

In order to determine the dual theory, we have to use the second equation in (\ref{YPEqs}) (the duality relation)
and solve for $Q$ in terms of $D$,
 \be\label{PthroughY}
  Q_{a|bc} \ = \ \partial^d D_{da|bc}
  -\frac{2}{D-2} \,\eta_{a[\underline{b}}\,\partial^dD_{de|}{}^{e}{}_{\underline{c}]}\;,
 \ee
which upon reinsertion into (\ref{firstorderY}) yields the dual action for $D$,
 \be\label{DualYAction}
  {\cal L} \ = \ \frac{1}{4}\,\partial_aD^{ab|cd}\,\partial^e D_{eb|cd}
  -\frac{1}{2(D-2)}\,\partial_aD^{ab|}{}_{b}{}^{c}\,\partial^d D_{de|}{}^{e}{}_{c}\;.
 \ee
Variation w.r.t.~$D$ yields the second-order equation
 \be\label{secondorderY}
  \partial_{[a}\partial^eD_{|e|b]|cd}-\frac{2}{D-2}\partial_{[a}\big(\eta_{b][c}\,
  \partial^eD_{ef|}{}^{f}{}_{d]}\big) \ = \ 0\,,
 \ee
which is equivalent to the result obtained from (\ref{PthroughY}) by taking a curl and
using the
Bianchi identity $\partial_{[a}Q_{b]|cd}=0$.

In the remainder of this section, we will analyze the dual theory in a little more detail.
We first decompose $D$ into its irreducible representations:
 \be\label{repr}
 \begin{split}
  D_{ab|cd}\; :\qquad   {\small \yng(1,1)}\;\otimes \; {\small \yng(1,1)}\hspace{0.2em}
  \ &= \ {\small \yng(1,1,1,1)} \;\oplus\; \widetilde{\small \yng(2,1,1)} \;\oplus \;
  \widetilde{\small \yng(2,2)} \;\oplus \;{\small \yng(1,1)}\;\oplus\; {\small \yng(2)}\;,
 \end{split}
 \ee
where we decomposed at the right-hand side into traceless tableaux (indicated by a tilde)
and the trace parts. Thus, the decomposition (into not yet irreducible representations) reads
 \be\label{Ydecompose}
  D_{ab|cd} \ = \ \tilde{D}_{ab|cd} + 4\,\eta_{[a[\underline{c}}\, \widehat{C}_{b]|\underline{d}]}\;,
 \ee
where $\tilde{D}$ is fully traceless, corresponding to the first three representations in (\ref{repr}),
and $\widehat{C}_{a|b}$ is a general 2-tensor (with antisymmetric and symmetric parts), corresponding
to the last two representations.

We will now show that the duality relations
imply the correct equations for the $(D-2,2)$ field. In order to simplify the index
manipulations we specialize to $D=4$, which shows already all essential features,
and for which the conventional dual to the $B$-field
is a scalar and the exotic dual is a $(2,2)$ tensor.
In this case we can decompose $D$ as
 \be\label{YD=4}
  D_{ab|cd} \ = \ \tfrac{1}{2}\epsilon_{ab}{}^{ef} B_{ef,cd}+4\eta_{[a[\underline{c}} C_{b]|\underline{d}]}
  -2\eta_{c[a }\eta_{b]d} C\;,
 \ee
where $B$ is the `Hodge-dual' form of the traceless $\tilde{D}$ in (\ref{Ydecompose}) and hence lives in the $(2,2)$ Young tableau. Moreover, we have redefined the general 2-tensor
for later convenience,
 \be
  {C}_{a|b} \ \equiv \ \widehat{C}_{a|b}-\tfrac{1}{2}\eta_{ab}\widehat{C}\;.
 \ee
The $\Sigma$ gauge symmetries can be decomposed as follows
 \be
  \Sigma_{abc|ef} \ \equiv \ \epsilon_{abc}{}^{d}\tilde{\Sigma}_{ef|d}\,, \qquad
  \tilde{\Sigma}_{ab|c}\,:\quad
   {\small \yng(1,1)}\;\otimes \; {\small \yng(1)}\hspace{0.2em} \ = \  {\small \yng(1,1,1)} \; \oplus  \;
    {\small \yng(2,1)}\;\;,
 \ee
so that we can write
 \be
  \tilde{\Sigma}_{ab|c} \ = \ \lambda_{ab,c}+\epsilon_{abcd}\,\xi^d\;,
 \ee
where $\lambda\in (2,1)$ and $\xi$ is  a new vector gauge parameter.
Applying the gauge transformations (\ref{YPgauge}) to (\ref{YD=4})
and using this decomposition of the gauge parameter one finds the following
gauge transformations for the component fields:
 \be\label{BandCgaugetransf}
  \begin{split}
   \delta B_{ab,cd} \ &= \ \partial_{[a}\lambda_{|cd|,b]}+\partial_{[c}\lambda_{|ab|,d]}\;, \\[0.5ex]
   \delta C_{a|b} \ &= \ 2\partial_{[a}\tilde{\xi}_{b]}-\partial_b\xi_a
   +\tfrac{1}{4}\epsilon_a{}^{cde}\partial_c\lambda_{de,b} \;.
  \end{split}
 \ee
The transformation in the first line is precisely the expected gauge transformation of a $(2,2)$
gauge field, c.f.~(\ref{D-22Gauge}), while the symmetry parametrized by $\mu$ in (\ref{D-22Gauge})
trivializes in $D=4$ because there is no $(1,2)$ Young  tableau.
Note that the extra field $C_{a|b}$ transforms under the gauge symmetry parametrized by
$\lambda_{ab,c}$.
The duality relation (\ref{PthroughY}) in terms of $B$ and ${C}$
reads
 \be\label{PGamma}
  Q_{b|cd} \ = \ -\tfrac{1}{3!}\epsilon_{b}{}^{efg}\, \Gamma_{efg,cd} \ + \ 2\partial_{[c} {C}_{b|d]}\;,
 \ee
with the generalized Christoffel symbol  (\ref{geneChris}).
It is an instructive exercise to verify the gauge invariance of this equation: Under `$b$-field gauge
transformations' with parameter $\tilde{\xi}_a$ the left- and right-hand sides are \textit{not} invariant, but their respective variations
precisely cancel. The right-hand side is manifestly invariant under the $\xi_a$ transformations, while
under $\lambda$ transformations the variations of the two terms on the right-hand side
cancel.

We next show that the duality relation implies as integrability condition
the desired field equation for the $(2,2)$ field. To this end we act on (\ref{PGamma})
with $\epsilon^{abij}\partial_a$, for which the left-hand side gives zero, and one obtains
 \be\label{eqGammaandCwithepsilon}
  0 \ = \ -\partial_a\Gamma^{aij}{}_{,cd}+2 \epsilon^{abij}\partial_a\partial_{[c}{C}_{b|d]}\;.
 \ee
Now summing over $i, c$ and $j, d$, the second term depending on $C$ drops out,
leaving
 \be\label{eqGammaandCwithepsilondos}
  0 \ = \  -\partial_a\Gamma^{acd}{}_{,cd} \ \equiv \ -\partial_{[a}\Gamma^{acd}{}_{,cd]} \qquad
  \Leftrightarrow \qquad  {\cal R}^{abc}{}_{,abc} \ = \ 0\;.
 \ee
Thus, we obtained the expected field equation (\ref{SCALAR}) for $D=4$, which
proves that the $(2,2)$ gauge field propagates
the single degree of freedom of the $b$-field in $D=4$.

\subsection{Dual action}
Let us finally determine and analyze the Lagrangian in terms of the dual fields, obtained by substituting (\ref{YD=4}) into (\ref{DualYAction}),
 \be
 \begin{split}
  {\cal L}[B,C] \ = \ &-\tfrac{1}{24}\,\Gamma^{abc,de}\, \Gamma_{abc,de} -\tfrac{1}{3!}\,\epsilon^{abcd}\,
  \Gamma_{bcd,}{}^{ef}\,\partial_{e}{C}_{a|f} \\[1ex]
  &+\tfrac{1}{2}\,\partial^a{C}^{b|c}\,\partial_a{C}_{b|c}
  -\tfrac{1}{2}\,\partial^c{C}^{a|b}\,\partial_b{C}_{a|c}
  -\tfrac{1}{2}\,\partial^c{C}^{a|b}\,\partial_a{C}_{c|b}\\[1ex]
  &+\partial_a{C}^{a|b}\,\partial_b{C}
  -\tfrac{1}{2}\,\partial^a{C}\,\partial_a{C}\;.
 \end{split}
 \ee
It is amusing to write this in a slightly more geometric form by defining the generalized
`Einstein tensor'
 \be
  {\cal G}_{a|b} \ \equiv \ \tfrac{1}{2}\big(-\square C_{a|b}+\partial^c\partial_a {C}_{c|b}+\partial^c\partial_b{C}_{a|c}
  -\partial_a\partial_b{C}+\eta_{ab}(\square {C}-\partial^c\partial^d{C}_{c|d})\big)\;,
 \ee
which satisfies the Bianchi identities 
 \be
  \partial^a{\cal G}_{a|b} \ = \ \partial^b{\cal G}_{a|b} \ = \ 0\;,
 \ee
and in terms of which the action reads
 \be\label{SimpleL}
    {\cal L} \ = \ -\tfrac{1}{24}\,\Gamma^{abc,de}\, \Gamma_{abc,de}  -\tfrac{1}{3!}\,\epsilon^{abcd}\,
  \Gamma_{bcd,}{}^{ef}\,\partial_{e}{C}_{a|f}  + C^{a|b}{\cal G}_{a|b}(C)\;.
 \ee
Note that decomposing $C$ into symmetric and antisymmetric parts, $C_{a|b}=s_{ab}+a_{ab}$,
with $s_{ab}\equiv s_{(ab)}$, $a_{ab}\equiv a_{[ab]}$,
the generalized Einstein tensor becomes
 \be
  {\cal G}_{a|b}(s,a) \ = \ G_{ab}(s)-\tfrac{1}{2}\,\partial^c h_{cab}(a)\;,
 \ee
in terms of the standard 3-form curvature $h_{abc}\equiv 3\partial_{[a}a_{bc]}$ and the
(linearized) Einstein tensor $G_{ab}=R_{ab} -\frac{1}{2}R\eta_{ab}$,
where
 \be
 \begin{split}
  R_{ab} \ \equiv \ \partial_a\gamma_{c}{}^{c}{}_{,b} -\partial^c \gamma_{ca,b}\;,\qquad 
  \gamma_{ab,c} \ \equiv \ \tfrac{1}{2}(\partial_a s_{bc}+\partial_bs_{ac}
  -\partial_cs_{ab})\;.
 \end{split}
 \ee
The  above Lagrangian can then be written as
\be
\begin{split}
    {\cal L} \ = \ &-\tfrac{1}{24}\,\Gamma^{abc,de}\, \Gamma_{abc,de} +\tfrac{1}{12}\,\epsilon^{abcd}\,
  \Gamma_{bcd,}{}^{ef}\,h_{aef}
  -\tfrac{1}{6}\,\epsilon^{abcd}\,\Gamma_{bcd,}{}^{ef}\,\gamma_{ae,f}
  \\[1ex]
  &+s^{ab} G_{ab}(s)+\tfrac{1}{6}\,h^{abc} h_{abc}
  \;.
  \end{split}
 \ee
Curiously, one obtains the conventional (linearized) Einstein-Hilbert term for $s_{ab}$ plus
the standard kinetic term for $a_{ab}$, both multiplied by an overall factor of $-2$.
These wrong-sing kinetic terms for a `graviton' and
a `Kalb-Ramond field' naively would lead one to
conclude that this theory propagates a ghost-like spin-2 mode and (in $D=4$) a scalar mode.
However, since the action is not diagonal and since these fields are subject to larger gauge
symmetries parameterized by $\lambda_{ab,c}$, there is no conflict with the equivalence to a single
scalar mode, which is guaranteed by the construction from a master action.

As a consistency check, let us verify
that this action indeed implies the expected field equation for the $(2,2)$ field.
Varying (\ref{SimpleL}) w.r.t.~$B_{ab,cd}$ and $C_{a|b}$, respectively,  yields
 \be\label{dualFieldEq}
  \begin{split}
   \partial^e\Gamma_{e\langle ab,cd\rangle} -  R^{\star}_{\langle ab,cd\rangle}(C)
   \ &= \ 0\;, \\[1ex]
   {\cal G}_{a|b}(C)+\tfrac{1}{12}\,\epsilon_{acde}\,\partial_f\Gamma^{cde,f}{}_{b} \ &= \ 0\;,
  \end{split}
 \ee
where $\langle\;\;\rangle$ denotes the projection onto the $(2,2)$ Young diagram
representation,\footnote{Explicitly, acting on a tensor $X_{ab|cd}$ that is antisymmetric in each index pair,
this projector reads
 \be
  X_{\langle ab|cd\rangle} \ \equiv \ \tfrac{1}{3}\big(X_{ab|cd} +X_{cd|ab} +\tfrac{1}{2} X_{ac|bd}
   -\tfrac{1}{2}X_{bc|ad} -\tfrac{1}{2} X_{ad|bc}
  +\tfrac{1}{2}X_{bd|ac} \big)\;.
 \ee}
and we defined the analogue of the linearized Riemann tensor for $C_{a|b}$ and its dualization
 \be
  R_{abcd}(C) \ \equiv \ 4\,\partial_{[ \underline{c}}\,\partial_{[a}\,C_{b]|\underline{d}]}\;, \qquad
  R^{\star}_{ab,cd}(C) \ \equiv \ \tfrac{1}{2}\,\epsilon_{ab}{}^{ef}\,R_{ef,cd}(C)\;.
 \ee
This Riemann tensor satisfies the Bianchi identity $R_{[abcd]}=0$, which in turn implies
that the double trace of $R^{\star}_{ab,cd}$ vanishes (note, however, that
$R_{[abc]d}$ generally is non-zero because $C$ carries an antisymmetric part).
As a consequence, taking the double trace of the first equation in
(\ref{dualFieldEq}), the $R^{\star}$ term drops out, implying the required
field equation  ${\cal R}^{abc}{}_{,abc}  =  0$, precisely as in (\ref{eqGammaandCwithepsilondos}).
The $(2,2)$ projection of the dual Riemann tensor in (\ref{dualFieldEq}) plays a role analogous
to the Weyl tensor in Einstein gravity (where it is left undetermined by the field
equations and hence encodes the propagating graviton degrees of freedom).
Here, on the contrary, the tensor $R^{\star}_{\langle ab,cd\rangle}$
is fully determined by the $(2,2)$ gauge potential,
in agreement with the non-propagating nature of $C_{a|b}$.



\section{Dualizations in Linearized DFT}
In this section we discuss the relations between dual and standard fields in Double Field Theory (DFT),
using  linearized DFT in the frame formulation
\cite{Siegel:1993th,Hohm:2010pp,Geissbuhler:2013uka}. We will add Lagrange multipliers (denoted as \emph{D}-fields in the following) to the linearized DFT action in order to enforce the Bianchi identities. This will allow us to obtain duality relations between the conventional fields and the $D$-fields and, as integrability conditions,  second order differential equations.

\subsection{Linearized DFT in frame formulation}
The fundamental fields in the frame formulation of DFT are the generalized vielbein ${E}_A{}^M$ and the generalized dilaton $d$. The vielbein transforms from the right under global $G=O(D,D)$ transformations and has a local $H=O(D-1,1)\times O(D-1,1)$ action from the left:
\begin{equation}
{E}'_A{}^{M}(X') \ = \ O^M{}_{N}\,{E}_B{}^N(X)\, h_A{}^B(X)  \;,~~X'^M \ = \ O^M{}_NX^N\;,
\end{equation}
where $O\in G$ and $h\in H$.
The generalized vielbein and the dilaton also transform under generalized coordinate transformations.
The frame field is subject to a covariant constraint, which
can be stated in terms of the `flattened' form of the $O(D,D)$ metric
 \be
  \eta_{MN} \ = \ \begin{pmatrix} 0 & 1\\ 1 & 0 \end{pmatrix}\;.
 \ee
In the original frame  formulation of DFT the subgroup $H=O(D-1,1)\times O(D-1,1)$ is embedded canonically,
indicated by the index split of the doubled Lorentz indices $A=(a,\bar{a})$, $a,\bar{a}=0,\ldots, D-1$, under which the flattened metric
is assumed to be diagonal,
 \be\label{FirstG}
  {\cal G}_{AB} \ \equiv \ {E}_{A}{}^{M} {E}_{B}{}^{N}\eta_{MN} \ \equiv \ 2 \,{\rm diag}(-\eta_{ab}, \eta_{\bar{a}\bar{b}}) \;,
 \ee
where $\eta_{ab}$ and $\eta_{\bar{a}\bar{b}}$ are two copies of the flat $D$-dimensional Lorentz metric
diag$(-+\cdots +)$,
and the relative sign between them is so that the overall signature is compatible with the $(D,D)$
signature of $\eta_{MN}$. 

A different but equivalent form of the constraint is given by choosing the
flattened metric so that it takes the same form as the $O(D,D)$ metric,
\begin{equation}\label{eqStandardParameterization}
\eta_{AB} \ \equiv \ \mathcal{E}_A{}^M\mathcal{E}_B{}^N\eta_{MN} \ = \ \begin{pmatrix}
0 & \delta^{a}{}_{b}\\
\delta_{a}{}^{b} & 0
\end{pmatrix},
\end{equation}
where we denoted the frame field by ${\cal E}_{A}{}^{M}$ to indicate that it satisfies a different constraint.
Due to this constraint, ${\cal E}_{A}{}^{M}$ is a proper $O(D,D)$ group element.
The flat indices split as $A=({}^a,{}_a)$
and, therefore, in this formalism
one has to carefully distinguish between upper and lower indices.
The tangent space indices are raised and lowered with $\eta_{AB}$ or ${\cal G}_{AB}$, depending
on the formalism.

The generalized metric encoding metric $g$ and $b$-field can be
defined conventionally in terms of the frame field.  For instance, in the formalism based on (\ref{eqStandardParameterization}), we define the $O(D-1,1)\times O(D-1,1)$ invariant metric
\begin{equation}\label{SMetric}
S_{AB} \ \equiv \ \begin{pmatrix}
\eta^{ab} & 0\\
0 & \eta_{ab}
\end{pmatrix},
\end{equation}
where $\eta_{ab}$ and $\eta^{ab}$ are again two copies of the flat Lorentz metric,
in terms of which the generalized metric can be written as
\begin{equation}
\mathcal{H}_{MN} \ = \ \mathcal{E}_M{}^A\, \mathcal{E}_N{}^B\, S_{AB}\,.
\end{equation}
In the following we use the perturbation theory for both formalisms,
with frame fields subject to either (\ref{FirstG}) or (\ref{eqStandardParameterization}),
because each is more convenient for different purposes.
In the remainder of this section we discuss the formalism based on (\ref{eqStandardParameterization}),
using the conventions of \cite{Geissbuhler:2013uka},
while the formalism based on (\ref{FirstG}) will be discussed and applied in sec.~5.

\medskip

We now discuss the frame-like perturbation theory,  
whose details have been developed in \cite{Hohm:2011dz} for flat and curved backgrounds. 
Here we consider perturbations around a constant background, 
writing
\begin{equation}
\mathcal{E}_A{}^M \ = \ \bar{\cal E}_A{}^{M}+ h_A{}^{B}\,\bar{\cal E}_{B}{}^{M}\;.
\end{equation}
The constraint (\ref{eqStandardParameterization}), which requires ${\cal E}_{A}{}^{M}$ to be
$O(D,D)$ valued, implies to first order in the fluctuation $h_{AB}+h_{BA}=0$.
We thus assume $h_{AB}$ to be antisymmetric.\;\footnote{Note, however, that beyond
first order this relation gets modified.}
Moreover, in the following we denote the linearization of the dilaton by $d$ and its
background value by $\bar{d}$.
The linearized theory is naturally written in terms of generalized coefficients of anholonomy,
also known as generalized fluxes \cite{Geissbuhler:2013uka,Geissbuhler:2011mx}, which are defined as
\begin{equation}\label{LinearizedFluxes}
\mathcal{F}_{ABC} \ = \ 3\, \mathcal{D}_{[A}h_{BC]}\;,
\qquad
\mathcal{F}_A \ = \ \mathcal{D}^{B}h_{BA}+2\,\mathcal{D}_Ad\;,
\end{equation}
with the flattened (doubled) derivative
\begin{equation}
\mathcal{D}_A \ \equiv \ \bar{\mathcal{E}}_{A}{}^M\partial_M\;.
\end{equation}
Note that in DFT we impose the `strong constraint' $\partial^MX\,\partial_MY=\partial^M\partial_MX=0$
for any fields $X,Y$,
which then implies ${\cal D}^A{\cal D}_A=0$ acting on arbitrary objects (which we will sometimes abbreviate
as ${\cal D}^2=0$).
It is then easy to verify that
the above coefficients of anholonomy  satisfy the Bianchi identities
\begin{equation}\label{LZ1}
\begin{split}
\mathcal{D}_{[A}\mathcal{F}_{BCD]} \ &= \ 0\;, \\[0.5ex]
\mathcal{D}^{C}\mathcal{F}_{CAB}+2\,\mathcal{D}_{[A}\mathcal{F}_{B]}  \ &= \ 0\;, \\[0.5ex]
\mathcal{D}^{A}\mathcal{F}_{A} \ &= \ 0\;.
\end{split}
\end{equation}
Conversely, it is straightforward to prove, using the Poincar\'e lemma and the strong constraint
${\cal D}^A{\cal D}_A=0$, that the general solution of these equations is given by (\ref{LinearizedFluxes}).

Let us now turn to the linearized DFT action, which takes the form
\begin{equation}\label{actionDFTflux}
S_{DFT} \ = \ \int d^{2D}X\, e^{-2\bar{d}}
\left(S^{AB}\mathcal{F}_A\mathcal{F}_B + \frac{1}{6}\,\breve{\mathcal{F}}^{ABC}\mathcal{F}_{ABC}\right)\;,
\end{equation}
where $\breve{\mathcal{F}}^{ABC}$ is defined as:
\begin{equation}
\breve{\mathcal{F}}^{ABC} \ \equiv \ \breve{S}^{ABCDEF}\mathcal{F}_{DEF}\, ,
\end{equation}
with the short-hand notation
\begin{equation}
\breve{S}^{ABCDEF}=\frac{1}{2}S^{AD}\eta^{BE}\eta^{CF}+\frac{1}{2}\eta^{AD}S^{BE}\eta^{CF}+\frac{1}{2}\eta^{AD}\eta^{BE}S^{CF}-\frac{1}{2}S^{AD}S^{BE}S^{CF}\;.
\end{equation}
The tensors $\breve{S}$ and $S$ satisfy the following identities:
\begin{equation}\label{InvolutiveProperty}
{\breve S}_{ABC}{}^{GHI}\breve{S}_{GHI}{}^{DEF} \ = \ \delta_{A}{}^D\delta_{B}{}^E\delta_{C}{}^F\;,
\qquad S_{A}{}^BS_{B}{}^{C}\ = \ \delta_{A}{}^C.
\end{equation}
The action \eqref{actionDFTflux} is invariant under infinitesimal generalized diffeomorphisms (with
the generalized coefficients of anholonomy being invariant to first order) and  local double Lorentz transformations
$\delta_{\Lambda}h_{AB}=\Lambda_{AB}$, with infinitesimal parameter $\Lambda_{AB}$ satisfying
 \be
  \Lambda_{AB}=-\Lambda_{BA}\;, \qquad
  S_A{}^C\Lambda_{CB}=S_B{}^C\Lambda_{AC}\;.
 \ee
In fact, the local Lorentz group leaves invariant the two metrics  (\ref{eqStandardParameterization})
and (\ref{SMetric}), which defines an $O(D-1,1)\times O(D-1,1)$ subgroup of $O(D,D)$. Under these
doubled Lorentz transformations, the coefficients of anholonomy transform as
\begin{equation}\label{LDLFABCandFA}
\delta_{\Lambda}\mathcal{F}_{ABC} \ = \ 3\,\mathcal{D}_{[A}\Lambda_{BC]}\;,
\qquad
\delta_{\Lambda}\mathcal{F}_A\ = \ \mathcal{D}^{B}\Lambda_{BA}\;.
\end{equation}
The equations of motion following from the linearized DFT action \eqref{actionDFTflux} for $h_{AB}$ and $d$,  respectively, are given by
\begin{equation}\label{LDFTEoME}
2\mathcal{D}^{[B}\mathcal{F}_{A}\,S^{C]A}+\mathcal{D}_{A}\breve{\mathcal{F}}^{ABC} \ = \ 0\;,
\end{equation}\\[-5ex]
\begin{equation}\label{LDFTEoMd}
2S^{AB}\mathcal{D}_B\mathcal{F}_A \ = \ 0\;.
\end{equation}

\subsection{Master action and duality relations}

We now pass to a first-order or master action as in previous sections, promoting
$\mathcal{F}_A$ and $\mathcal{F}_{ABC}$ to independent fields and
introducing (totally antisymmetric) Lagrange multipliers $D^{ABCD}$, $D^{AB}$ and $D$ that
enforce the Bianchi identities.
The action thus reads
\begin{equation}\label{DFTfirstorderaction}
\begin{split}
S \ = \ \int dX \,e^{-2\bar{d}}&\,\bigg[S^{AB}\mathcal{F}_A\mathcal{F}_B
+\frac{1}{6}\,\breve{\mathcal{F}}^{ABC}\mathcal{F}_{ABC}\\
&+D^{ABCD}\, \mathcal{D}_{A}\mathcal{F}_{BCD} 
+D^{AB}\big(\mathcal{D}^{C}\mathcal{F}_{CAB}+2\,\mathcal{D}_{A}\mathcal{F}_{B}\big)
+D\,\mathcal{D}^{A}\mathcal{F}_A \bigg]\;.
\end{split}
\end{equation}
Varying w.r.t.~the fundamental fields $D^{ABCD}$, $D^{AB}$, $D$, $\mathcal{F}_{ABC}$
and $\mathcal{F}_A$, respectively, we obtain the field equations
\begin{equation}\label{eqdeltaDABCD}
\mathcal{D}_{[A}\mathcal{F}_{BCD]} \ = \ 0\;,
\end{equation}\\[-5ex]
\begin{equation}\label{eqdeltaDAB}
\mathcal{D}^{C}\mathcal{F}_{CAB}+2\,\mathcal{D}_{[A}\mathcal{F}_{B]} \ = \ 0\;,
\end{equation}\\[-5ex]
\begin{equation}\label{eqdeltaD}
\mathcal{D}^{A}\mathcal{F}_A \ = \ 0\;,
\end{equation}\\[-5ex]
\begin{equation}\label{DualityRelationFABC}
\begin{split}
\breve{\mathcal{F}}^{ABC}  \ = \ 3\,\big( \mathcal{D}_DD^{DABC}+\mathcal{D}^{[A}D^{BC]}\big)\;,
\end{split}
\end{equation}\\[-5ex]
\begin{equation}\label{DualityRelationFA}
2S^{AB}\mathcal{F}_B-2\,\mathcal{D}_BD^{BA}-\mathcal{D}^{A}D \ = \ 0\;.
\end{equation}
With the first three equations we recover the Bianchi identities, which can be
solved as in (\ref{LinearizedFluxes}), giving back the original (linearized) DFT.
The last two equations (\ref{DualityRelationFABC}) and (\ref{DualityRelationFA})
can then be interpreted as the duality relations. From these we may obtain the original
second-order linearized DFT equations as integrability conditions.
To this end, we act on eq.~(\ref{DualityRelationFABC}) with $\mathcal{D}_A$ and obtain
\begin{equation}\label{eqcurlDualityRelationFABC}
\mathcal{D}_A\breve{\mathcal{F}}^{ABC} \ = \ -2\,{\cal D}^{[B}{\cal D}_{A}D^{|A|C]}\;,
\end{equation}
where we have used $\mathcal{D}_{[A}\mathcal{D}_{B]}=0$ and the strong constraint $\mathcal{D}^2=0$.
Now we can use (\ref{DualityRelationFA}) in order to eliminate ${\cal D}_A D^{AC}$ on the right-hand
side, which gives back the linearized field equation (\ref{LDFTEoME}).
Similarly, by acting on eq.~(\ref{DualityRelationFA}) with $\mathcal{D}_A$ and using the strong constraint
one obtains the linearized dilaton equation of motion (\ref{LDFTEoMd}).

Let us now discuss the gauge symmetries in the dual formulation.
First, the duality relations and master action are invariant under the following gauge transformations:
\begin{equation}\label{SigmaTransformations}
\begin{split}
\delta D^{ABCD} \ &= \ \mathcal{D}_E\Sigma{}^{EABCD}+\mathcal{D}^{[A}\Sigma^{BCD]}\;,\\
\delta D^{AB} \ &=  \
\mathcal{D}^{[A}\Sigma^{B]}+\frac{3}{4}\,\mathcal{D}_E\Sigma{}^{EAB}\;,\\
\delta D \ &= \ \mathcal{D}_A\Sigma^A\;,
\end{split}
\end{equation}
where $\Sigma^{ABCDE}=\Sigma^{[ABCDE]}$ and $\Sigma^{ABC}=\Sigma^{[ABC]}$.
The $D$-fields also transform under double Lorentz transformations. Using (\ref{LDLFABCandFA})
in the above duality relations, one finds
\begin{equation}
\begin{split}
\delta_\Lambda D_{ABCD} \ = \ 0\;,\hskip 1truecm
\delta_\Lambda D_{AB} \ = \ -S^E{}_{[A}\Lambda_{B]E}\;,\hskip 1truecm
\delta_\Lambda D \ = \ 0\;.
\end{split}
\end{equation}

\subsection{Dual DFT}

 Let us now investigate the equations of motion for the theory in terms of the dual
 $D$-fields. These are obtained from the Bianchi identities (\ref{LZ1}) and
 the duality relations (\ref{DualityRelationFABC})--(\ref{DualityRelationFA}). First, we need to solve
 the duality relations for the coefficients of anholonomy in terms of the dual $D$-fields,
 which yields, using eq.~(\ref{InvolutiveProperty}),
\begin{subequations}\label{DualityRelationFABCFA}
\begin{empheq}{align}
\mathcal{F}_{ABC} \ &= \ 3\,\breve{S}_{ABC}{}^{DEF}\big( \mathcal{D}^{G}D_{GDEF}
+\mathcal{D}_{[D}D_{EF]}\big)\;,\\[1ex]
\mathcal{F}_A & \ = \ S_A{}^B\big(\mathcal{D}^{C}D_{CB}+\frac{1}{2}\,\mathcal{D}_BD\big)\;.
\end{empheq}
\end{subequations}
Inserting these into the Bianchi identities (\ref{LZ1}), we obtain
\begin{align}\label{2ndDeq}
0 \ &= \ \breve{S}_{[ABC|EFG|}\,\mathcal{D}_{D]}\Big(\mathcal{D}_HD^{HEFG}+\mathcal{D}^{[E}D^{FG]} \Big)\;,\\
0 \ &= \ 3\,\breve{S}_{CABDEF}\mathcal{D}^{C}\Big(\mathcal{D}_GD^{GDEF}+\mathcal{D}^{[D}D^{EF]}\Big)
+2\,S_{C[B}\mathcal{D}_{A]}\Big(\mathcal{D}_DD^{DC}
+ \frac{1}{2}\,\mathcal{D}^{C}D\Big)\;,\\\label{2ndDeqtres}
0 \ &= \ S_{AB}\mathcal{D}^{A}\mathcal{D}_{C}D^{CB}+\frac{1}{2}\,S^{AB}\mathcal{D}_{A}\mathcal{D}_{B}D\;.
\end{align}

In order to illuminate further these equations for the dual $D$-fields,
let us introduce the following field strengths:
\begin{equation}\label{GABCstrength}
G^{ABC} \ \equiv \ 3\,\left(\mathcal{D}_DD^{DABC} +\mathcal{D}^{[A}D^{BC]}\right),
\end{equation}
and
\begin{equation}\label{GAstrength}
G^{A} \ \equiv \ \mathcal{D}_B D^{BA}+\frac{1}{2}\,\mathcal{D}^{A}D\;,
\end{equation}
which are invariant under the $\Sigma$-transformations (\ref{SigmaTransformations}). In terms of these field strengths the duality relations take the following simpler form:
\begin{equation}\label{DRFwithG}
\begin{split}
\mathcal{F}^{ABC} \ & = \ \breve{S}^{ABCDEF}\,G_{DEF}, \\
\mathcal{F}^{A} \ & = \ S^{AB}\, G_B\;. \\
\end{split}
\end{equation}
Finally, defining ${\cal G}_{ABC}\equiv \breve{S}_{ABC}{}^{DEF}G_{DEF}$ and ${\cal G}_A\equiv S_A{}^BG_B$,
the second-order equations (\ref{2ndDeq})--(\ref{2ndDeqtres}) for the dual fields take exactly the same form as
the Bianchi identities for the original fields. Our final form of duality relations between fluxes and dual fluxes is then
\begin{equation}\label{DRFwithcalG}
\begin{split}
\mathcal{F}_{ABC} \ & = \ \mathcal{G}_{ABC}\ , \\
\mathcal{F}_{A} \ & = \ \mathcal{G}_{A}\ . \\
\end{split}
\end{equation}
The set of equations for the original and dual system is summarized in Table \ref{BoxEoMandBI}.

\begin{table}[h]
\hspace{0.1cm}
\begin{tabular}{ |c|c|c| }
\cline{2-3}
\multicolumn{1}{c|}{} & \multirow{2}{*}{DFT} & \multirow{2}{*}{Dual DFT} \\
\multicolumn{1}{c|}{} & & \\[0.2cm]
 \hline
 &  - - - & $\mathcal{D}_{[D}{\cal G}_{ABC]}=0$\\[0.2cm]
E.o.M's &  $2\mathcal{D}^{[B}\mathcal{F}_{A}\,S^{C]A}+\mathcal{D}_{A}\breve{\mathcal{F}}^{ABC}=0$ &
$\mathcal{D}^{C}{\cal G}_{CAB}+ 2\mathcal{D}_{[A}{\cal G}_{B]} =0$ \\[0.2cm]
 &  $2S^{AB}\mathcal{D}_A\mathcal{F}_B=0$ & $\mathcal{D}^{A}{\cal G}_A=0$ \\[0.2cm]
  \hline
 &  $\mathcal{D}_{[D}\mathcal{F}_{ABC]}=0$ & - - - \\[0.2cm]
B.I's & $\mathcal{D}^{C}\mathcal{F}_{CAB}+2\mathcal{D}_{[A}\mathcal{F}_{B]}=0$ & $2\mathcal{D}^{[A}{\cal G}_CS^{B]C}+\mathcal{D}_C\breve{{\cal G}}^{CAB}=0$ \\[0.2cm]
 &  $\mathcal{D}^{A}\mathcal{F}_{A}=0$ & $2S^{AB}\mathcal{D}_{A}{\cal G}_B=0$ \\[0.2cm]
  \hline
\end{tabular}
\caption{\footnotesize Comparison of equations of motion and Bianchi identities between DFT and dual DFT. \label{BoxEoMandBI}}
\end{table}
\vskip .5truecm


 \section{Geometric form of dual DFT action}
 In this section we elaborate on the geometric form of the dual DFT action.
 We first present a master action in terms of connections that, in a sense, is complementary to
 that presented in sec.~4, but which leads to equivalent results.
 Finally, we determine the dual action and write it in a geometric form that is
 completely analogous to the dual action for the exotic duals discussed in sec.~3.

\subsection{DFT action in connection form }
In order to define the master action in a (semi-)geometric form,
let us first review the linearized frame-like geometry of DFT, based on a
frame field $E_{A}{}^{M}$, where the flat indices split as $A=(a,\bar{a})$.
Since the frame field is subject to (\ref{FirstG}), expansion about  a constant background,
 \be
  E_{A}{}^{M} \ = \ \bar{E}_{A}{}^{M}-h_{A}{}^{B}\bar{E}_{B}{}^{M}\;,
 \ee
leads to the following first-order constraints on the fluctuations
 \be
  h_{a\bar{b}} \ = \ -h_{\bar{b}a}\;, \qquad h_{ab} \ \equiv \ h_{[ab]}\;, \qquad h_{\bar{a}\bar{b}} \ \equiv \
   h_{[\bar{a}\bar{b}]}\;.
 \ee
The first field is physical, encoding the symmetric metric  fluctuation and the
antisymmetric $b$-field fluctuation. The final two fields are pure gauge w.r.t.~the local $O(D-1,1)\times O(D-1,1)$ tangent space symmetry.   Indeed, defining
$\partial_A\equiv \vev{E_{A}{}^{M}}\partial_M$, the linearized gauge transformations can be written as
 \be
  \delta h_{AB} \ = \ \partial_A\xi_{B}-\partial_B\xi_{A}+\Lambda_{AB}\;,
 \ee
where $\Lambda_{AB}={\rm diag}(\Lambda_{ab},\Lambda_{\bar{a}\bar{b}})$, and therefore
 \be\label{localh}
  \begin{split}
   \delta h_{a\bar{b}} \ &= \ \partial_a\xi_{\bar{b}}-\partial_{\bar{b}}\xi_{a}\;, \\
   \delta h_{ab} \ &= \ 2\,\partial_{[a}\,\xi_{b]}+\Lambda_{ab}\;, \\
   \delta h_{\bar{a}\bar{b}} \ &= \ 2\,\partial_{[\bar{a}}\, \xi_{\bar{b}]}+\Lambda_{\bar{a}\bar{b}}\;,
  \end{split}
 \ee
while the dilaton transforms as
 \be
  \delta d \ = \ -\tfrac{1}{2}(\partial_{a}\xi^{a}+\partial_{\bar{a}}\xi^{\bar{a}})\;.
 \ee
From (\ref{localh}) we infer that $h_{ab}$ and $h_{\bar{a}\bar{b}}$ can be
gauged away.
The spin connection components of the linearized theory
read
 \be\label{DFTconn}
 \begin{split}
  \omega_{a\bar{b}\bar{c}} \ &= \ -2\,\partial_{[\bar{b}}h_{|a|\bar{c}]}
  +\partial_ah_{\bar{b}\bar{c}}\;, \\
  \omega_{\bar{a}bc} \  &= \ \;\; 2\,\partial_{[b}h_{c]\bar{a}}+\partial_{\bar{a}}h_{bc}\;, \\
  \omega_{a} \ &\equiv \ \omega_{ba}{}^{b} \ = \ \partial^bh_{ab}+\partial^{\bar{b}}h_{a\bar{b}}
  +2\partial_a d\;, \\
  \omega_{\bar{a}} \ &\equiv \ \omega_{\bar{b}\bar{a}}{}^{\bar{b}} \ = \ -\partial^b h_{b\bar{a}}
  +\partial^{\bar{b}}h_{\bar{a}\bar{b}}+2\partial_{\bar{a}}d \;, \\
  \omega_{[abc]} \ &= \ \partial_{[a}h_{bc]} \;, \\
  \omega_{[\bar{a}\bar{b}\bar{c}]} \ &= \ \partial_{[\bar{a}}h_{\bar{b}\bar{c}]}
  \;.
 \end{split}
 \ee
These objects indeed transform as connections for the doubled local Lorentz symmetry:
 \be\label{genlocalLor}
  \begin{split}
   \delta \omega_{a\bar{b}\bar{c}} \ &= \ \partial_{a}\Lambda_{\bar{b}\bar{c}}\;, \qquad
   \delta \omega_{\bar{a}bc} \ = \ \partial_{\bar{a}}\Lambda_{bc}\;, \\
   \delta \omega_{a} \ &= \ \partial^b \Lambda_{ab}\;, \qquad \;\;\,
   \delta \omega_{\bar{a}} \ = \ \partial^{\bar{b}}\Lambda_{\bar{a}\bar{b}}\;.
  \end{split}
 \ee
In particular, the connections are fully invariant under generalized diffeomorphisms.
The above connections satisfy the Bianchi identities
 \be\label{connBianchi}
 \begin{split}
   \partial_a\omega^a+\partial_{\bar{a}}\omega^{\bar{a}} \ &= \ 0\;, \\[1ex]
  \partial^{\bar{a}} \omega_{\bar{a}bc} - 2\,\partial_{[b}\, \omega_{c]} +3\,\partial^a\omega_{[abc]} \ &= \  0\;, \\[1ex]
  \partial^a\omega_{a\bar{b}\bar{c}} -2\,\partial_{[\bar{b}}\,\omega_{\bar{c}]}+3\,\partial^{\bar{a}}
  \omega_{[\bar{a}\bar{b}\bar{c}]} \ &= \  0  \;, \\[1ex]
  \partial^c \omega_{\bar{b}ac}-\partial^{\bar{c}}\omega_{a\bar{b}\bar{c}}+\partial_a\omega_{\bar{b}}
  -\partial_{\bar{b}}\omega_{a}\ &= \  0  \;, \\[1ex]
   \partial_{[a}\,\omega_{|\bar{d}|bc]} - \partial_{\bar{d}}\,\omega_{[abc]} \ &= \ 0\;, \\[1ex]
  \partial_{[\bar{a}}\,\omega_{|d|\bar{b}\bar{c}]} - \partial_{d}\,\omega_{[\bar{a}\bar{b}\bar{c}]} \ &= \ 0\;,  \\[1ex]
  \partial_{[a}\,\omega_{b]\bar{c}\bar{d}}+\partial_{[\bar{c}}\,\omega_{\bar{d}\,]ab} \ &= \ 0\;, \\[1ex]
   \partial_{[a}\,\omega_{bcd]} \ &= \ 0\;, \\[1ex]
  \partial_{[\bar{a}}\, \omega_{\,\bar{b}\bar{c}\bar{d}\,]} \ &= \ 0
  \;.
 \end{split}
 \ee
This is a rather extensive list of identities, but except for the first one they are
all consequences of the algebraic Bianchi identity for the full Riemann tensor,
${\cal R}_{[ABC]D}=0$,  see \cite{Hohm:2012mf}, and are
also equivalent to (\ref{LZ1}).

We now give invariant curvatures in order to define the dynamics of linearized DFT.
There is a linear generalized Riemann tensor,
 \be\label{DFTRiemann}
  {\cal R}_{ab,\bar{c}\bar{d}} \ \equiv \ \partial_{[a}\,\omega_{b]\bar{c}\bar{d}}-\partial_{[\bar{c}}\,\omega_{\bar{d}]ab}
  \ = \ -4\,\partial_{[a}\partial_{[\bar{c}}\,h_{b]\bar{d}]}\;,
 \ee
which, however, does not have a non-linear completion.
The linearized (generalized) Ricci tensor (which is \textit{not} the trace of
the above Riemann tensor) reads
 \be\label{Ricci}
  {\cal R}_{a\bar{b}} \ \equiv \ -\partial^c\omega_{\,\bar{b}ac}+\partial_{\bar{b}}\,\omega_{a}
  \ \equiv \ -\partial^{\bar{c}}\omega_{a\bar{b}\bar{c}}+\partial_{a}\omega_{\bar{b}}\;,
 \ee
where the equivalence of the two definitions follows from the fourth Bianchi identity in (\ref{connBianchi}).
The explicit expression in components reads
 \be
  {\cal R}_{a\bar{b}} \ = \ \square h_{a\bar{b}}-\partial_{a}\partial^c h_{c\bar{b}}
  +\partial_{\bar{b}}\partial^{\bar{c}} h_{a\bar{c}}+2\partial_{a}\partial_{\bar{b}}d \;,
 \ee
where $\square\equiv \partial^a\partial_a\equiv -\partial^{\bar{a}}\partial_{\bar{a}}$.
As it should be, the pure gauge degrees of freedom dropped out.
Also note that there are differential Bianchi identities relating (\ref{DFTRiemann}) to (\ref{Ricci}),
 \be\label{diffBianchi}
  \partial^{\bar{c}}{\cal R}_{ab,\bar{c}\bar{d}} \ = \ -2\,\partial_{[a} {\cal R}_{b]\bar{d}}\;, \qquad
  \partial^a {\cal R}_{ab,\bar{c}\bar{d}} \ = \ 2\,\partial_{[\bar{c}}{\cal R}_{|b|\bar{d}]}\;.
 \ee
The linearized scalar curvature is
 \be
  {\cal R} \ \equiv \ -\partial^a\omega_a
  \ \equiv \ \partial^{\bar{a}}\omega_{\bar{a}} \ = \ -2\square d -\partial^{a}\partial^{\bar{b}} h_{a\bar{b}} \;,
 \ee
where we have given the explicit component expression in the last step.
Finally, the linearized DFT action in terms of the connections reads
 \be\label{DFT2}
 \begin{split}
  {\cal L}^{(2)}_{\rm DFT} \ = \ \frac{1}{2}\big(\,& \omega^{a\bar{b}\bar{c}}\omega_{a\bar{b}\bar{c}}
  +3\,\omega^{[\bar{a}\bar{b}\bar{c}]}\omega_{[\bar{a}\bar{b}\bar{c}]}+2\,\omega^{\bar{a}}\omega_{\bar{a}}\,\\
  &-\omega^{\bar{a}bc}\omega_{\bar{a}bc}-3\,\omega^{[abc]}\omega_{[abc]}-2\,\omega^a\omega_a
  \big)\;,
 \end{split}
 \ee
whose general variation reads $\delta{\cal L}=4\delta h^{a\bar{b}}\,{\cal R}_{a\bar{b}}-8\delta d\,{\cal R}$.
Let us note that, upon inserting (\ref{DFTconn}),
the two lines in the above action actually give the same result, by virtue of the strong constraint
and the relative sign between them, but for our present purposes this action is convenient
because it treats barred and unbarred indices on the same footing.

\subsection{Master action}
We now give a first-order master action that can be used to define the dual theory and
in which the connections are promoted to independent fields, in analogy to previous sections.
Apart from that, the approach is complementary to that used in previous sections in that the
dual fields do not enter the master action as Lagrange multipliers but rather emerge upon `solving' the field equations
by reinterpreting them as Bianchi identities. This approach is of course fully equivalent to that
used before (the difference being whether the fields or their duals enter the master action that
serves as the starting point),
but it is reassuring to confirm explicitly that both procedures give the same result.

We now treat the connections as independent fields and replace the linearized DFT action (\ref{DFT2})
by the first-order action
 \be\label{MasterACTion}
 \begin{split}
  {\cal L}^{(1)}_{\rm DFT} \ = \ &-\tfrac{1}{2} \omega^{a\bar{b}\bar{c}}\omega_{a\bar{b}\bar{c}}
  +\omega^{a\bar{b}\bar{c}}\big(-2\partial_{\bar{b}}h_{a\bar{c}}+\partial_a h_{\bar{b}\bar{c}}\big)
  -\tfrac{3}{2}\,\omega^{[\bar{a}\bar{b}\bar{c}]}\omega_{[\bar{a}\bar{b}\bar{c}]}
  +3\,\,\omega^{[\bar{a}\bar{b}\bar{c}]} \partial_{\bar{a}}h_{\bar{b}\bar{c}} \\[0.5ex]
  &-\omega^{\bar{a}}\omega_{\bar{a}}  +2\,\omega^{\bar{a}}(-\partial^b h_{b\bar{a}} +\partial^{\bar{b}}h_{\bar{a}\bar{b}}
  +2\partial_{\bar{a}}d )\\[0.5ex]
  &+\tfrac{1}{2}\omega^{\bar{a}bc}\omega_{\bar{a}bc}-\omega^{\bar{a}bc}
  \big(2\partial_bh_{c\bar{a}}+\partial_{\bar{a}}h_{bc}\big)
  +\tfrac{3}{2}\,\omega^{[abc]}\omega_{[abc]}-3\,\omega^{[abc]}\partial_{a}h_{bc}\\[0.5ex]
  &+\omega^a\omega_a-2\,\omega^{a}\big(\partial^{\bar{b}}h_{a\bar{b}}+\partial^b h_{ab} +2\partial_ad\big)\;.
 \end{split}
 \ee
The field equations for the $\omega$ determine them in terms of the physical fields as given in (\ref{DFTconn}),
so that
reinserting into the action we recover (\ref{DFT2}). On the other hand, varying with respect to
$d$, $h_{a\bar{b}}$, $h_{ab}$ and $h_{\bar{a}\bar{b}}$, respectively, we obtain
 \be\label{MasterEQUATIONS}
  \begin{split}
    \partial_{\bar{a}}\omega^{\bar{a}}-\partial_a\omega^a \ &= \ 0\;,  \\[0.5ex]
   -\partial_{\bar{c}}\,\omega^{a\bar{b}\bar{c}}-\partial_{c}\,\omega^{\bar{b}ac}+\partial^a\omega^{\bar{b}}
   +\partial^{\bar{b}}\omega^{a} \ &= \ 0\;, \\[0.5ex]
    \partial_{\bar{a}}\omega^{\bar{a}bc} +3\,\partial_a\omega^{[abc]}-2\,\partial^{[b}\,\omega^{c]}    \ &= \ 0\;,  \\[0.5ex]\
    \partial_a\omega^{a\bar{b}\bar{c}}+3\,\partial_{\bar{a}}\omega^{[\bar{a}\bar{b}\bar{c}]}-2\,\partial^{[\bar{b}}\,
    \omega^{\bar{c}]} \ &= \ 0\;.
  \end{split}
 \ee
Expressing $\omega$ in terms of the physical fields, the first two equations give the DFT equations
${\cal R}_{a\bar{b}}=0$ and ${\cal R}=0$, while the last two equations are the second and third Bianchi identity in
(\ref{connBianchi}).

In order to determine the dual theory we interpret now all four of the equations (\ref{MasterEQUATIONS})  as
Bianchi identities and solve them in terms of dual fields.
We proceed hierarchically, starting with the first equation, which can be solved as
 \be\label{traceSOL}
 \begin{split}
  \omega^{\bar{a}} \ &= \ \partial_{\bar{b}}D^{\bar{a}\bar{b}}+\partial_bD^{b\bar{a}}+\partial^{\bar{a}}D\;, \\[0.5ex]
  \omega^{a} \ &= \ \partial_{b}D^{ab}+\partial_{\bar{b}}D^{a\bar{b}}-\partial^aD\;,
 \end{split}
 \ee
with $D^{\bar{a}\bar{b}}$ and $D^{ab}$ antisymmetric, $D^{b\bar{a}}$ unconstrained and a singlet $D$.
This result can be obtained as follows.
First, the non-singlet terms follow from the standard Poincar\'e lemma, writing the equation as
$\partial_A\Omega^A=0$ for $\Omega^A\equiv (-\omega^a,\omega^{\bar{a}})$, which
implies $\Omega^A = \partial_BD^{AB}$ for antisymmetric $D^{AB}$, whose components
give the above $D$ fields. The only subtlety is that the derivatives are subject to
the strong constraint, which allows for the singlet term that drops out by
$\partial^{\bar{a}}\partial_{\bar{a}}=-\partial^a\partial_a$.
Thus, (\ref{traceSOL}) is the general solution of the first equation in (\ref{MasterEQUATIONS}).

Next, we turn to the second equation in (\ref{MasterEQUATIONS}),
where we can eliminate $\omega^a$ and $\omega^{\bar{a}}$
according to (\ref{traceSOL}). We first solve the equation for the special case that all these $D$ fields  are
zero:
 \be
   -\partial_{\bar{c}}\,\omega^{a\bar{b}\bar{c}}-\partial_{c}\,\omega^{\bar{b}ac} \ = \ 0\;.
 \ee
This is solved by
 \be\label{Dsolone}
 \begin{split}
  \omega^{a\bar{b}\bar{c}} \ &= \ \partial_{\bar{d}}D^{\bar{b}\bar{c}\bar{d},a}
  +\partial_d D^{da,\bar{b}\bar{c}} \;, \\[0.5ex]
  \omega^{\bar{b}ac} \ &= \ \partial_dD^{cda,\bar{b}}+\partial_{\bar{d}}D^{ac,\bar{b}\bar{d}}\;,
 \end{split}
 \ee
where the $D$ fields are antisymmetric in each group of similar indices.
Including now the trace connections we need to solve the inhomogeneous equation
 \be
  \partial_{\bar{c}}\,\omega^{a\bar{b}\bar{c}}+\partial_{c}\,\omega^{\bar{b}ac} \ = \
  \partial^a\partial_{\bar{c}}D^{\bar{b}\bar{c}}+\partial^a\partial_cD^{c\bar{b}}
  +\partial^{\bar{b}}\partial_{c}D^{ac}+\partial^{\bar{b}}\partial_{\bar{c}}D^{a\bar{c}}\;,
 \ee
where we note that the singlet $D$ dropped out.
This equation is solved by
 \be\label{inhomoD}
 \begin{split}
  \omega^{a\bar{b}\bar{c}} \ &= \ \partial^aD^{\bar{b}\bar{c}}+2\,\partial^{[\bar{b}} D^{|a|\bar{c}]}\;, \\[0.5ex]
  \omega^{\bar{b}ac} \ &= \ \partial^{\bar{b}}D^{ac} + 2\,\partial^{[a} D^{c]\bar{b}}\;,
 \end{split}
 \ee
which can be verified by employing the strong constraint again.
Thus, the general solution is given by the sum of (\ref{Dsolone}) and (\ref{inhomoD}),
 \be\label{omega3D}
  \begin{split}
   \omega^{a\bar{b}\bar{c}} \ &= \ \partial_{\bar{d}}D^{\bar{b}\bar{c}\bar{d},a}
  +\partial_d D^{da,\bar{b}\bar{c}}+\partial^aD^{\bar{b}\bar{c}}+2\,\partial^{[\bar{b}} D^{|a|\bar{c}]} \;, \\[0.5ex]
  \omega^{\bar{b}ac} \ &= \ \partial_dD^{cda,\bar{b}}+\partial_{\bar{d}}D^{ac,\bar{b}\bar{d}}
  +\partial^{\bar{b}}D^{ac} + 2\,\partial^{[a} D^{c]\bar{b}} \;.
 \end{split}
 \ee

Finally, we solve the last two equations in (\ref{MasterEQUATIONS}).  Inserting (\ref{traceSOL}) and
(\ref{omega3D}) determines $\omega^{[abc]}$ up to solutions of $\partial_a\omega^{[abc]}=0$,
which by the Poincar\'e lemma are given by $\partial_dD^{abcd}$ for a new totally antisymmetric tensor
$D_{abcd}$. Applying the same reasoning to
$\omega^{[\bar{a}\bar{b}\bar{c}]}$ introduces the new field $D_{\bar{a}\bar{b}\bar{c}\bar{d}}$, and we finally find for the connections in terms of the dual fields,
  \be\label{edDualityrel}
  \begin{split}
   \omega^{a\bar{b}\bar{c}} \ &= \ \partial_{\bar{d}}D^{\bar{b}\bar{c}\bar{d},a}
  +\partial_d D^{da,\bar{b}\bar{c}}+\partial^aD^{\bar{b}\bar{c}}+2\,\partial^{[\bar{b}} D^{|a|\bar{c}]} \;, \\[0.5ex]
   \omega^{\bar{a}bc} \ &= \ \partial_dD^{bcd,\bar{a}}+\partial_{\bar{d}}D^{bc,\bar{a}\bar{d}}
  +\partial^{\bar{a}}D^{bc} + 2\,\partial^{[b} D^{c]\bar{a}} \;, \\[0.5ex]
  \omega^{\bar{a}} \ &= \ \partial_{\bar{b}}D^{\bar{a}\bar{b}}+\partial_bD^{b\bar{a}}+\partial^{\bar{a}}D\;, \\[0.5ex]
  \omega^{a} \ &= \ \partial_{b}D^{ab}+\partial_{\bar{b}}D^{a\bar{b}}-\partial^aD\;, \\[0.5ex]
    \omega_{[abc]} \ &= \ \partial_{[a}D_{bc]}-\partial^d D_{abcd} -\tfrac{1}{3}\,\partial^{\bar{d}}D_{abc,\bar{d}}
  \;, \\[0.5ex]
   \omega_{[\bar{a}\bar{b}\bar{c}]} \ &= \
   \partial_{[\bar{a}}D_{\bar{b}\bar{c}]}-\partial^{\bar{d}}D_{\bar{a}\bar{b}\bar{c}\bar{d}}-\tfrac{1}{3}\,\partial^d
   D_{\bar{a}\bar{b}\bar{c},d}
  \;.
 \end{split}
 \ee
For the reader's convenience we summarize here the dual $D$ fields:
 \be
  \begin{split}
   &D\;, \qquad D_{ab}\;, \qquad D_{\bar{a}\bar{b}}\;, \qquad D_{a\bar{b}}\;, \\[1ex]
   &D_{abc\bar{d}}\;, \quad D_{\bar{a}\bar{b}\bar{c}d}\;,\quad D_{ab\bar{c}\bar{d}}\;, \quad D_{abcd}\;, \quad
   D_{\bar{a}\bar{b}\bar{c}\bar{d}}\;.
  \end{split}
 \ee
Comparing with  the list of Bianchi identities (\ref{connBianchi}) we infer that the $D$ fields
and Bianchi identities are in one-to-one correspondence.
Thus, these fields could be used as Lagrange multipliers to impose the Bianchi identities,
confirming the equivalence with the master action procedure discussed in sec.~4.

We now turn to the dual gauge symmetries that leave (\ref{edDualityrel}) invariant and thus describe the redundancies between the $D$ fields.
For the two-index fields one finds
 \be\label{twoindexD}
  \begin{split}
   \delta_{\Sigma}D_{ab} \ &= \ 2\,\partial_{[a} \Sigma_{b]}+\partial^c\Sigma_{abc}+\partial^{\bar{c}}\Sigma_{ab,\bar{c}}\;,
   \\[0.5ex]
   \delta_{\Sigma}D_{\bar{a}\bar{b}} \ &= \
   -2\,\partial_{[\bar{a}} \Sigma_{\bar{b}]}+\partial^{\bar{c}}\Sigma_{\bar{a}\bar{b}\bar{c}}
   +\partial^{c}\Sigma_{\bar{a}\bar{b},c}\;, \\[0.5ex]
   \delta_{\Sigma}D_{a\bar{b}} \ &= \  \partial_{a}\Sigma_{\bar{b}}-\partial_{\bar{b}}\Sigma_{a}
   +\partial^{c}\Sigma_{ca,\bar{b}}+\partial^{\bar{c}}\Sigma_{\bar{c}\bar{b},a}\;, \\[0.5ex]
   \delta_{\Sigma}D \ &= \ \partial_a\Sigma^a+\partial_{\bar{a}}\Sigma^{\bar{a}}\;.
  \end{split}
 \ee
Note that the dual diffeomorphism parameters $\Sigma_a$ and $\Sigma_{\bar{a}}$ act
on these fields in exactly the same way as the original diffeomorphism parameters
$\xi_a$ and $\xi_{\bar{a}}$ act on $h_{ab}$, $h_{\bar{a}\bar{b}}$, $h_{a\bar{b}}$ and $d$.
For the four-index field we find
 \be\label{DUALDIff}
  \begin{split}
   \delta_{\Sigma} D^{\bar{a}\bar{b}\bar{c},d} \ &= \ \partial_{\bar{e}}\Sigma^{\bar{a}\bar{b}\bar{c}\bar{e},d}
   -\partial^d\Sigma^{\bar{a}\bar{b}\bar{c}}+3\, \partial^{[\bar{a}}\,\Sigma^{\bar{b}\bar{c}],d}\;, \\[0.5ex]
   \delta_{\Sigma}D^{ab,\bar{c}\bar{d}} \ &= \
   2\,\partial^{[a}\,\Sigma^{|\bar{c}\bar{d}|,b]}-2\,\partial^{[\bar{c}}\,\Sigma^{|ab|,\bar{d}]}\;, \\[0.5ex]
   \delta_{\Sigma}D^{abc,\bar{d}} \ &= \ \partial_e\Sigma^{abce,\bar{d}}
   -\partial^{\bar{d}}\Sigma^{abc}+3\,\partial^{[a}\Sigma^{bc],\bar{d}} \;, \\[0.5ex]
   \delta_{\Sigma}D^{abcd} \ &= \ \partial_{e}\Sigma^{abcde}+ \tfrac{4}{3}\,\partial^{[a}\Sigma^{bcd]}
   -\tfrac{1}{3}\,\partial_{\bar{e}}\Sigma^{abcd,\bar{e}}\;, \\[0.5ex]
     \delta_{\Sigma}D^{\bar{a}\bar{b}\bar{c}\bar{d}} \ &= \
     \partial_{\bar{e}}\Sigma^{\bar{a}\bar{b}\bar{c}\bar{d}\bar{e}}+
      \tfrac{4}{3}\,\partial^{[\bar{a}}\Sigma^{\bar{b}\bar{c}\bar{d}]}
   -\tfrac{1}{3}\,\partial_{{e}}\Sigma^{\bar{a}\bar{b}\bar{c}\bar{d},{e}}\;.
  \end{split}
 \ee
It can be verified by a straightforward computation that these transformations
leave  (\ref{edDualityrel}) invariant. Finally, in order for (\ref{edDualityrel}) to transform under local
Lorentz transformations as required by (\ref{genlocalLor}), the
$D$ fields need to transform as
 \be\label{LorentzonD}
  \delta_{\Lambda} D^{ab} \ = \ \Lambda^{ab}\;, \qquad
  \delta_{\Lambda} D^{\bar{a}\bar{b}} \ = \ \Lambda^{\bar{a}\bar{b}}\;.
 \ee
Thus, exactly as for  $h_{ab}$ and $h_{\bar{a}\bar{b}}$, these fields are pure gauge.

\subsection{Geometric action for dual DFT fields}
Let us now insert (\ref{edDualityrel})
into the master action (\ref{MasterACTion}) in order to obtain the action for the dual $D$ fields.
The terms involving the original fields drop out because these fields  enter linearly, multiplying constraints that have been solved in terms of the $D$ fields. The second-order action therefore reads
 \be\label{DFT22}
 \begin{split}
  {\cal L}^{(2)}_{\rm DFT} \ = \ -\frac{1}{2}\big(\,& \omega^{a\bar{b}\bar{c}}\omega_{a\bar{b}\bar{c}}
  +3\,\omega^{[\bar{a}\bar{b}\bar{c}]}\omega_{[\bar{a}\bar{b}\bar{c}]}+2\,\omega^{\bar{a}}\omega_{\bar{a}}\,\\
  &-\omega^{\bar{a}bc}\omega_{\bar{a}bc}-3\,\omega^{[abc]}\omega_{[abc]}-2\,\omega^a\omega_a
  \big)\;,
 \end{split}
 \ee
with the connections given by (\ref{edDualityrel}).
This takes precisely the same form as (\ref{DFT2}), except that the overall sign has changed.
The computation of inserting (\ref{edDualityrel}) is simplified by using that the dependence of $\omega$
on $D_{a\bar{b}}$, $D_{ab}$, $D_{\bar{a}\bar{b}}$ and $D$ is precisely analogous
to the expressions in terms of the original fields, up to the following identifications,
 \be
  D_{a\bar{b}}\;\;\rightarrow\;\; -h_{a\bar{b}}\;, \quad D_{ab}\;\;\rightarrow\;\;  -h_{ab}\;, \quad
  D_{\bar{a}\bar{b}}\;\;\rightarrow\;\;  h_{\bar{a}\bar{b}}\;, \quad D \;\;\rightarrow\;\; 2d\;,
 \ee
and an overall sign for the connections with unbarred Lie algebra indices,
which is irrelevant since the connections enter the action quadratically.
A direct computation yields the explicit form of the dual Lagrangian,
 \be\label{DFTDualAction}
  \begin{split}
   {\cal L} \ = \ &-\tfrac{1}{2}\,\partial_{\bar{d}}D^{\bar{b}\bar{c}\bar{d},a}\,\partial^{\bar{e}}D_{\bar{b}\bar{c}\bar{e},a}
   -\partial_{\bar{d}}D^{\bar{b}\bar{c}\bar{d},a}\,\partial^{e} D_{ea,\bar{b}\bar{c}}
   -\tfrac{1}{2}\,\partial_d D^{da,\bar{b}\bar{c}}\,\partial^e D_{ea,\bar{b}\bar{c}} \\[1ex]
   &+\tfrac{1}{2}\,\partial_d D^{bcd,\bar{a}}\,\partial^e D_{bce,\bar{a}}+\partial_d D^{bcd,\bar{a}}\,\partial^{\bar{e}}
   D_{bc,\bar{a}\bar{e}}
   +\tfrac{1}{2}\, \partial_{\bar{d}} D^{bc,\bar{a}\bar{d}}\,\partial^{\bar{e}} D_{bc,\bar{a}\bar{e}}\\[1ex]
   &-\tfrac{3}{2}\,\partial_{\bar{d}}D^{\bar{a}\bar{b}\bar{c}\bar{d}}\,\partial^{\bar{e}}D_{\bar{a}\bar{b}\bar{c}\bar{e}}
   -\partial_{\bar{d}} D^{\bar{a}\bar{b}\bar{c}\bar{d}}\,\partial^eD_{\bar{a}\bar{b}\bar{c},e}
   -\tfrac{1}{6}\,\partial_d D^{\bar{a}\bar{b}\bar{c},d}\,\partial^{e}D_{\bar{a}\bar{b}\bar{c},e}
   \\[1ex]
   &+\tfrac{3}{2}\,\partial_d D^{abcd}\,\partial^e D_{abce} +\partial_dD^{abcd}\,\partial^{\bar{e}} D_{abc,\bar{e}}
   +\tfrac{1}{6}\,\partial_{\bar{d}} D^{abc,\bar{d}}\,\partial^{\bar{e}}D_{abc,\bar{e}}
   \\[1ex]
   &-D^{ab,\bar{c}\bar{d}}\,{\cal R}_{ab,\bar{c}\bar{d}}(D_{a\bar{b}})
   -{\cal L}^{(2)}_{\rm DFT} (D_{a\bar{b}},D)\;.
  \end{split}
 \ee
Note that in the last line we encounter the standard
linearized DFT Lagrangian ${\cal L}^{(2)}$, but for $D_{a\bar{b}}$ and $D$, with the `wrong' overall sign,
in complete analogy to the mixed
Young tableau action discussed in sec.~3. Also in perfect analogy to that discussion is that this wrong-sign
kinetic term does not indicate the presence of ghosts, for the action is not diagonal.
Rather, the off-diagonal term is proportional to  the linearized Riemann tensor (\ref{DFTRiemann}),
but expressed in terms of $D_{a\bar{b}}$.
Thus, the $\Sigma_{a}$ and $\Sigma_{\bar{a}}$ transformations are manifest symmetries of this action,
while the invariance under the remaining dual diffeomorphisms (\ref{DUALDIff}) can be verified
by a direct computation.
Also note that the fields $D_{\bar{a}\bar{b}}$ and $D_{ab}$ dropped out, as it should be in view of the St\"uckelberg-type
Lorentz invariance (\ref{LorentzonD}).

We close this section by discussing two of the $D$-field equations, because
they exhibit an intriguing structure. Varying w.r.t.~$D_{a\bar{b}}$ and $D$ we obtain
 \be
  {\cal R}_{a\bar{b}}(D) \ = \ \partial^c\partial^{\bar{d}} D_{ac,\bar{b}\bar{d}}\;, \qquad {\cal R}(D) \ = \ 0\;.
 \ee
For the first equation neither the left-hand side nor the right-hand side are dual diffeomorphism invariant
under transformations with parameter $\Sigma_{ab,\bar{c}}$ and $\Sigma_{\bar{a}\bar{b},c}$,
but their variations precisely cancel against each other.
The field equation for $D_{ab,\bar{c}\bar{d}}$ reads
\be\label{S=REQ}
     {\cal R}_{ab\bar{c}\bar{d}} \ = \  S_{ab\bar{c}\bar{d}} \;,
 \ee
where we defined
 \be
  S_{ab\bar{c}\bar{d}} \ \equiv  \  \partial_{[a}\partial^e D_{|e|b],\bar{c}\bar{d}}
  +\partial_{[\bar{c}}\,\partial^{\bar{e}} D_{|ab|,\bar{d}]\bar{e}}
  +\partial_{[a}\partial^{\bar{e}} D_{|\bar{c}\bar{d}\bar{e}|,b]}
  +\partial_{[\bar{c}}\,\partial^e D_{|abe|,\bar{d}]}\;.
 \ee
Thus, intriguingly, the equation takes the form of a second-order duality relation, relating
the (linearized) Riemann tensor to a `dual' Riemann tensor. As above, both sides are not separately
invariant under dual diffeomorphisms with parameter $\Sigma_{ab,\bar{c}}$ and $\Sigma_{\bar{a}\bar{b},c}$,
but the full equation of course is, as it should be and as may be verified by a quick computation.


\section{Comparison of results}

In Section 4 we have shown that at the linearized level the DFT equations and Bianchi identities for the fluxes  $\mathcal{F}_{ABC}$ and  $\mathcal{F}_{A}$
arise from first order duality equations given, for instance, in eq.~(\ref{DRFwithcalG}), relating these fluxes to the dual fluxes $\mathcal{G}_{ABC}$ and $\mathcal{G}_{A}$. The dual fluxes are defined in terms of the field strengths $G_{ABC}$ of the dual potentials (the $D$-fields) in eqs.~(\ref{GABCstrength}) and (\ref{GAstrength}). The field equations and Bianchi identities for the fields and the dual fields are listed in Table \ref{BoxEoMandBI}.
The aim of this section is to show that if one restricts all DFT fields to only depend on $x$, {\it i.e.}~if one sets
$\tilde{\partial}^\mu\Phi =0$ for any DFT field $\Phi$, one recovers the previous results of dualization: the standard dualities between
the 2-form and the $(D-4)$-form and  between the graviton (plus dilaton) and the mixed-symmetry $(D-3,1)$ potential discussed in sec.~2, 
and the exotic duality between the $2$-form and the mixed-symmetry $(D-2,2)$ potential  discussed in sec.~3.

The dual potentials introduced in sec.~4 are $D_{ABCD}$, $D_{AB}$ and $D$. Upon breaking $O(D-1,1) \times O(D-1,1)$ to the diagonal subgroup, the field $D_{ABCD}$ can be decomposed as
\begin{equation}
D_{ABCD} \; \rightarrow \; D^{abcd} \quad D^{abc}{}_d \quad D^{ab}{}_{cd} \quad D^a{}_{bcd} \quad D_{abcd}\;,
\end{equation}
while the field $D_{AB}$ decomposes as
\begin{equation}
D_{AB} \; \rightarrow \; D^{ab} \quad D^{a}{}_b  \quad D_{ab}\;.
\end{equation}
When reducing to $x$-space we use, by a slight abuse of notation, the same symbols for the components of the DFT $D$-fields and the supergravity $D$-fields. The identification uses the ordering of the indices as given above to match the results of the previous sections. The same applies for the components of $G_{ABC}$. We make an exception,
in the following subsection, for the identification of the components of $D_{AB}$ and $D$ with the ones in $x$-space:
\begin{equation}\label{primeconventions}
\begin{split}
&D^{ab} \ \rightarrow \ D'^{ab}\,,\\
&D^{a}{}_{b} \ \rightarrow \ D'^{a}{}_{b}\,,~~~~~~~D \ \rightarrow \  D''\;, \\
&D_{ab} \ \rightarrow \ D'_{ab}\;,\\
\end{split}
\end{equation}
the convention being that $x$-space fields carrying a prime can be gauged or redefined away.

If one inserts the above identifications into eq.~(\ref{DFTfirstorderaction}), one recovers the first order actions of Section 2. In particular, the fields $D^{abcd}$, $D^{abc}{}_d$ and $D^{\prime ab}$ are precisely the potentials that we introduced in Section 2 when we performed the standard dualization for the 2-form and the graviton plus dilaton system. 
This requires that, in $x$-space, the fields $D^{\prime a}{}_b$ and $D^{\prime\prime}$ can be redefined away 
and/or are irrelevant for the analysis. 
We will also see that $D^{a}{}_{bcd}$, $D_{abcd}$ and $D_{ab}$ trivialize in $x$-space frame.

One can also recover the duality relations for each field by performing the decomposition directly in the duality relation (\ref{DRFwithcalG}). We first identify the components of $\mathcal{F}_{ABC}$ in $x$-space as:
\begin{equation}\label{FLuxes}
\begin{split}
\mathcal{F}_{ABC} \ &= \ \{H_{abc},f_{ab}{}^c,Q_a{}^{bc},R^{abc}\}\;,\\
\mathcal{F}_{A} \ &= \ \{f_{a},Q^{a}\}\;,
\end{split}
\end{equation}
which at this stage are just labels for the components of the ${\cal F}$ flux. As we will see,
$H_{abc}$, $f_{ab}{}^c$ and $f_a$ play the same role as in in Section 2, and we will discuss later $Q_a{}^{bc}$, $Q^a$ and $R^{abc}$, which are related to non-geometric fluxes. Note that because of the presence of the tensor ${\breve S}_{ABCDEF}$ in the definition of $\mathcal{G}_{ABC}$ in terms of $G_{ABC}$, eq.~(\ref{DRFwithcalG}) relates a given component of $\mathcal{F}_{ABC}$ to different components of  ${G}_{ABC}$ and thereby to different components of the dual potentials. This has to be understood as follows: if one turns on a particular component of the flux  $\mathcal{F}_{ABC}$, eq.~(\ref{DRFwithcalG}) still gives equations for all  the dual potentials. The equations for the dual potentials dual to the vanishing fluxes will furnish algebraic relations among the different components of ${G}_{ABC}$, and  after reinserting these relations into the duality relation for the non-vanishing fluxes one finds that this is dual to a specific component of ${G}_{ABC}$ suitably antisymmetrized. This will also be discussed in each case in the remainder of this section, which is organized as follows. In the first subsection we will show how from DFT one recovers the standard dualizations of Section 2, while in the second subsection we will show how the exotic dualization of Section 3 is also contained in DFT. Finally, in the third subsection we will briefly discuss the remaining dual fields,
which are related to non-geometric fluxes such as the $R$-flux.

\subsection{Standard duality relations for the 2-form and graviton plus dilaton}

The truncation of the action given in  eq.~(\ref{DFTfirstorderaction}) to $x$-space with only either the $H$-flux or the $f$-flux turned on straightforwardly reproduces the field theory analysis of Section 2. In the case of the $H$-flux, only the component $D^{abcd}$ of $D_{ABCD}$ appears in the action, and one immediately recovers
eq.~(\ref{masteractionHtheory}). In the case of the $f$-flux, one turns on only the component $D^{abc}{}_d$ in $D_{ABCD}$ and $D^{\prime ab}$ in $D_{AB}$ to recover precisely the action in eq.~(\ref{masteractiongravanddil}). The analysis performed in Section 2 showed that $D^{\prime ab}$ is pure gauge while $D^{abc}{}_d$ describes both the dual of the graviton and the dual of the dilaton.

As anticipated at the beginning of this section, a more careful analysis is required if one wants to perform the truncation at the level of the duality relations. In the case of the
$H$-flux, the duality relation (\ref{DRFwithcalG}) simply gives $H_{abc} = \mathcal{G}_{abc}$, with the other components of $\mathcal{G}_{ABC}$ vanishing. In terms of $G_{ABC}$ this gives
 \begin{equation}
 H_{abc} =\mathcal{G}_{abc}	 = \frac{1}{2}\eta_{cf}G_{ab}{}^f-\frac{1}{2}\eta_{be}G_{ac}{}^e+\frac{1}{2}\eta_{ad}G_{bc}{}^d-\frac{1}{2}\eta_{ad}\eta_{be}\eta_{cf}G^{def} \ \ .\label{extractiontildeGfromDFTdos}
 \end{equation}
In this equation both $G^{abc}$ and $G_{ab}{}^c$ occur, but one has to take into account also the equation for the vanishing dual flux $\mathcal{G}_{a}{}^{bc}$, which gives
\begin{equation}
0=\mathcal{G}_a{}^{bc} = \frac{1}{2}\eta^{be}G_{ae}{}^c - \frac{1}{2}\eta^{cf}G_{af}{}^b - \frac{1}{2}\eta_{ad}\eta^{be}\eta^{cf}G_{ef}{}^d+\frac{1}{2}\eta_{ad}G^{dbc}\;,  \label{extractiontildeGfromDFTuno}
\end{equation}
implying the algebraic relation
  \begin{equation}
G_{ab}{}^c=-\eta_{be}\eta_{ad}G^{dec} \;.
\end{equation}
Upon inserting this relation into eq.~(\ref{extractiontildeGfromDFTdos}) one obtains
\begin{equation}
\mathcal{G}_{abc}=-2\eta_{ad}\eta_{be}\eta_{cf}G^{def} \; ,
\end{equation}
which is in agreement with (\ref{tildeHintermsofGabc}), identifying $\mathcal{G}_{abc}=\tilde{H}_{abc}$ and using the definition  of $G^{abc}$ given in eq.~(\ref{GABCstrength}).

We now perform the same analysis for the graviton-dilaton system. Turning on only the fluxes $f_{ab}{}^c$ and $f_a$ in eq.~(\ref{DRFwithcalG}) we must recover eq. (\ref{dualityrelationintermsofgabc}),
where $\mathcal{G}_{ab}{}^c$ is identified with $g_{ab}{}^c$ and $\mathcal{G}_{a}$ with $g_a$. In terms of $G_{ABC}$, one has
\begin{equation}
\mathcal{G}_{ab}{}^c  =\frac{1}{2}\eta^{cf}G_{abf}-\frac{1}{2}\eta_{ad}\eta_{be}\eta^{cf}G_f{}^{de}-\frac{1}{2}\eta_{ad}G_b{}^{dc}+\frac{1}{2}\eta_{be}G_a{}^{ec} \; . \label{calGvsGdualgravanddil}
\end{equation}
The two components $G_{abc}$ and $G_a{}^{bc}$ that occur in this equation are related by the condition that the dual flux $\mathcal{G}^{abc}$ vanishes, which yields the relation
\begin{equation}
0  = -\frac{1}{2}\eta^{ad}\eta^{be}\eta^{cf}G_{def}+\frac{1}{2}\eta^{cf}G_f{}^{ab}-\frac{1}{2}\eta^{be}G_e{}^{ac}+\frac{1}{2}\eta^{ad}G_d{}^{bc}\;.
\end{equation}
Inserting this into eq.~(\ref{calGvsGdualgravanddil}) one  obtains
\begin{equation}\label{matchDFTandcompogravplusdil}
\mathcal{G}_{ab}{}^c = \eta_{be}G_{a}{}^{ec}-\eta_{ad}G_{b}{}^{dc} \;,
\end{equation}
which precisely reproduces eq.~(\ref{dualityrelationintermsofgabc}) by using (\ref{GABCstrength}). It is also straightforward to show that $\mathcal{G}_a$ coincides with $g_a$ defined in (\ref{dualityrelationintermsofgabc}) after using eq.~(\ref{GAstrength}).

\subsection{$Q$-flux dualization from DFT}

We now consider the truncation to $x$-space of the DFT dualization for the $Q$-flux component in (\ref{FLuxes}) and show that it reproduces the exotic dualization of the 2-form discussed in Section~3.
We start from the  first order action (\ref{DFTfirstorderaction}), specialized to the $Q$-flux components,
and reduce to $x$-space,
\begin{align}\nonumber
S[Q,D] \ = \  \int {\rm d}^Dx\, \Big(&Q^{a}Q_{a} -\frac{1}{4}Q_a{}^{bc}Q^a{}_{bc}-\frac{1}{2}Q_{a}{}^{bc}Q_{b}{}^a{}_c
\\
&+ 3D^{ab}{}_{cd} \,\partial_a Q_b{}^{cd} + 2D^{a}{}_b \big( \partial_c Q_a{}^{bc} + \partial_a Q^b \big)
+ D\, \partial_a Q^a \Big) \ , \label{lagrangemultforQflux}
\end{align}
where the fields $D^{ab}{}_{cd}\equiv D^{[ab]}{}_{[cd]}$, $D^{ a}{}_b $, $D$ and
$Q^a$, $Q_a{}^{bc}$ are independent, and we dropped the primes relative to (\ref{primeconventions}).
The field equations for the $D$-fields read
\begin{eqnarray}
& &\partial_{[a} Q_{b]}{}^{cd} \ = \ 0 \,,\nonumber \\[0,5ex]
& & \partial_c Q_a{}^{bc} + \partial_a Q^b \ =  \ 0\,, \nonumber \\[0.5ex]
& & \partial_a Q^a \ = \ 0 \;,  \label{BianchiforQandcalFDFT}
\end{eqnarray}
which are the  Bianchi identities (\ref{LZ1}), reduced to $x$-space and
specialized to the components $Q_a{}^{bc}$ and $Q^a$.
The solution of these equations is
\begin{equation}\label{QSOL}
Q_a{}^{bc} \ = \ \partial_a \beta^{bc}\;,  \qquad Q^a \ = \ \partial_b\beta^{ba} + \textrm{constant} \;,
\end{equation}
and we will see in the following that the constant term is irrelevant.  
The field equations for $Q^a$ and $Q_a{}^{bc}$ yield the duality relations
\begin{equation}\begin{split}
2Q_a \ & = \ 2\partial_bD^b{}_a+\partial_a D\;,\\
-\frac{1}{2}Q^a{}_{bc} -  \frac{1}{2}Q_b{}^a{}_c+\frac{1}{2}Q_c{}^a{}_b \ & = \ 3\partial_eD^{ea}{}_{bc}
-2\partial_{[b}D^a{}_{c]}\;,\\
\end{split}\end{equation}
which are equivalent to the duality relations following from (\ref{DualityRelationFABC}) and (\ref{DualityRelationFA})
upon specializing to the $Q$-fluxes.

Comparing with the master action discussed in sec.~3, we observe that here we have Lagrange multiplier
fields, $D_a{}^b$ and $D$, which have no analogues in that previous analysis, but we will now show that
these fields are irrelevant. We first note that (\ref{lagrangemultforQflux}) is invariant under the
gauge transformations with local parameter $\chi$
 \be
  \delta_{\chi}D \ = \ \chi\;, \qquad \delta_{\chi}D^{a}{}_{b} \ = \ -\tfrac{1}{2}\,\chi\,\delta^a{}_{b}\;, \qquad
  \delta_{\chi}D^{ab}{}_{cd} \ = \ -\tfrac{1}{3}\,\chi\,\delta^{[a}{}_{c}\,\delta^{b]}{}_{d}\;,
 \ee
with $\delta_{\chi}Q=0$. These act as a St\"uckelberg symmetry on $D$. Thus, we can gauge this field
to zero.\footnote{Note that this gauge invariance cannot be realized in the $O(D,D)$ covariant formalism of DFT,
for it acts on the trace part of $D^a{}_b$ and the double trace part of $D^{ab}{}_{cd}$. There are no analogous
traces of the covariant and fully antisymmetric fields $D_{AB}$ and $D_{ABCD}$.}
Equivalently, we can express the action directly in terms of the gauge invariant objects
 \be
 \begin{split}
  \widehat{D}^a{}_{b} \ \equiv \ D^a{}_b \ + \  \tfrac{1}{2}\,D\,\delta^a{}_{b} \;, 
  \qquad \widehat{D}^{ab}{}_{cd} \ \equiv \ D^{ab}{}_{cd} \ + \ \tfrac{1}{3}\,D\, \delta^{[a}{}_{c}\,\delta^{b]}{}_{d}\;,
 \end{split}
 \ee
which yields
\begin{align}\nonumber\label{Qaction2}
S[Q,D] \ = \  \int {\rm d}^Dx\, \Big(&Q^{a}Q_{a} -\frac{1}{4}Q_a{}^{bc}Q^a{}_{bc}-\frac{1}{2}Q_{a}{}^{bc}Q_{b}{}^a{}_c
\\
&+ 3\widehat{D}^{ab}{}_{cd} \,\partial_a Q_b{}^{cd} + 2\widehat{D}^{a}{}_b \big( \partial_c Q_a{}^{bc} + \partial_a Q^b \big)
 \Big) \ .
\end{align}
As expected, the singlet $D$ field dropped out. The field $\widehat{D}^{a}{}_b$ cannot be eliminated similarly by a
gauge symmetry. 
Rather, its own field equation yields the second of the Bianchi identites in (\ref{BianchiforQandcalFDFT}), 
and back-substituting their solution (\ref{QSOL}) into the action (\ref{Qaction2}) gives the free Kalb-Ramond action
for the $b$-field, which at the linearized level is equivalent to the `$\beta$-supergravity' for the bi-vector field
$\beta^{ab}\equiv b^{ab}$ (with the indices raised by the flat Minkowski metric) \cite{Andriot:2012an}. 
Note, in particular, that the constant term in (\ref{QSOL}) contributes to the Lagrangian only an irrelevant 
constant and a total derivative term.  Therefore, it is physically equivalent to set the constant to zero, 
in which case $Q^a = Q_b{}^{ba}$ and the second  and third Bianchi identity are no longer independent but are traces of the first one.
Thus, on-shell the above action is equivalent to the same action with $Q^a = Q_b{}^{ba}$ and with the only
Lagrange multiplier being $\widehat{D}^{ab}{}_{cd}$, enforcing the first Bianchi identity in (\ref{BianchiforQandcalFDFT}).
This action is then manifestly equivalent to the master action (\ref{firstorderY}) discussed
in sec.~3.\footnote{Note that the third $Q^2$ term in (\ref{Qaction2}) is absent in (\ref{firstorderY}),
but upon eliminating $Q$ both actions agree up to total derivatives, which is sufficient for the
equivalence as master actions.}
Thus, we have shown that in the $Q$-flux sector the DFT dualization reduces to the
exotic dualization of the $B$-field into a mixed-symmetry potential with a  $(D-2,2)$ Young tableau.

\subsection{The $R$-flux}
We now consider the $R$-flux contribution of (\ref{FLuxes})  in the
truncation of the master action (\ref{DFTfirstorderaction}) to $x$-space. The action reduces to
  \begin{equation}
  S \ = \ \int {\rm d}^D x   \big( D^a{}_{bcd}\, \partial_a R^{bcd} + {D}'_{ab}\, \partial_c R^{cab}\big) \; , \label{dualRfluxmultipliers}
  \end{equation}
where $D^a{}_{bcd}=D^a{}_{[bcd]}$ and $D'_{ab}=D'_{[ab]}$.
Note that the field ${D}'_{ab}$ can be absorbed into the trace of $D^a{}_{bcd}$.
The equations for the dual potentials in this case simply imply that $R^{abc}$ has to be constant and hence that in this sector the fields carry no degrees of freedom. This is consistent with the form of the $R$-flux in $x$-space at the non-linear level:
\begin{equation}
R^{abc} \ = \ 3\,\beta^{[a|e|}\partial_e\beta^{bc]}\;,
\end{equation}
whose linearization vanishes for vanishing $\beta$ background.
The duality then implies that the dual flux $\mathcal{G}^{abc}$ also vanishes.

Finally, let us also note that the field $D_{abcd}$ disappears from the action in $x$-space since it couples to a Bianchi identity for the $R$-flux that explicitly contains a derivative $\tilde{\partial}^\mu$ with respect to the dual coordinate. The field $D_{abcd}$ can be written as a $(10,4)$ gauge field in $D=10$ by using the epsilon tensor, as can be deduced by writing its gauge transformation from
eq.~(\ref{SigmaTransformations}) and keeping only $x$ derivatives.  On the other hand, in an $O(D,D)$ frame in which we take all the fields to depend only on the coordinates $\tilde{x}$, $D_{abcd}$ would become the `standard' dual of the field $\beta$ since the $R$-flux takes the form
\begin{equation}
R^{abc} \ = \ 3\,\tilde{\partial}^{[a}\beta^{bc]}\;,
\end{equation}
which plays precisely the same role as the $H$-flux in $x$-space.
An analogous inversion of roles also holds for all other fields, as is guaranteed by the $O(D,D)$ invariance of the action (\ref{DFTfirstorderaction}). We summarize this in Table~\ref{tabla2}:
\begin{table}[h]
\begin{center}
\begin{tabular}{ccc|ccc}
& $x$-space  & & &$\tilde{x}$-space &\\
\hline
& & & & & \\
$b_{ab}$ & $\leftrightarrow$ & $D^{abcd}$ & $\beta^{ab}$ & $\leftrightarrow$ & $D_{abcd}$\\
$h_{a|}{}^{b}$ & $\leftrightarrow$ & $D^{abc}{}_{d}$ & $h^{a|}{}_{b}$ & $\leftrightarrow$ & $D^{a}{}_{bcd}$\\
$\beta^{ab}$ & $\leftrightarrow$ & $D^{ab}{}_{cd}$ & $b_{ab}$ & $\leftrightarrow$ & $D^{ab}{}_{cd}$ \\
\end{tabular}
\caption{Dual fields for the Kalb-Ramond field, vielbein fluctuation and $\beta$-field in $x$ and $\tilde{x}$-space.}
\label{tabla2}
\end{center}
\end{table}


\section{Conclusions and Outlook}

In this paper we have determined the dualization of double field theory (dual DFT) at the linearized level, which
captures in addition to the conventional dual fields in $D=10$ string theory (the 6-form dual to the Kalb-Ramond 2-form
and the $8$-form dual to the dilaton)
fields in mixed-Young tableaux representations, such as the dual of the graviton and an exotic dual
of the 2-form, plus additional fields. The dual fields can be organized into a
totally antisymmetric 4-tensor under the T-duality group $O(D,D)$, as suggested by previous studies,
but it turns out that
defining an $O(D,D)$ covariant master action (and, consequently, an action
for the dual fields) requires extra fields.

A careful analysis shows, however, that reducing the
dual DFT to the physical spacetime yields precisely the expected dual theories.
In particular, we analyzed the exotic dualization of the 2-form, following
the strategy introduced in \cite{Boulanger:2012df}, which is illuminating because it shows
that, besides the dual $(D-2,2)$ gauge potential, extra fields are needed that
carry the representations and gauge symmetries of metric and 2-form fluctuations. Consequently,
they
enter the action with (linearized) Einstein-Hilbert and Kalb-Ramond terms, but due to non-trivial couplings
to the dual fields this does not upset the
counting of degrees of freedom.
Similarly, the dual DFT carries, besides the 4th rank $O(D,D)$ tensor,
fields with the same representations and gauge symmetries as in the original DFT and, therefore, they enter the
action with the usual (linearized) generalized curvature scalar of DFT. Again, because of the coupling
to the dual fields,
this does not indicate the presence of unphysical modes, as is also guaranteed by the master action.

This unusual feature may provide important pointers for the full non-linear theory yet to be constructed.
In general, there are strong no-go theorems implying, under rather mild assumptions, that there is no
non-linear action  for a mixed-Young tableau field that is invariant under a
deformation of the linear gauge symmetries \cite{Bekaert:2002uh,Bekaert:2004dz}.
However, in the $O(D,D)$  covariant framework analyzed here, this problem presents itself
in a quite different fashion. Because of the coupling to extra fields (carrying the representations
of the original DFT fields), the no-go theorem is not applicable, and hence it may well be that there is a
consistent non-linear deformation of the dual DFT action (\ref{DFTDualAction}).
For instance, this would require finding a non-linear extension of the field equation (\ref{S=REQ}).
The linearized DFT Riemann tensor appearing on the left-hand side of that equation by itself does not
have a non-linear extension \cite{Hohm:2012mf} (which in turn is the reason that higher-derivative
$\alpha'$-corrections require a deformation of the framework \cite{Hohm:2013jaa}), but
it is natural to speculate that a non-linear extension exists which deforms not only the
left-hand side but also the right-hand side that encodes the mixed Young tableau fields.\footnote{Intriguingly, 
one may thus speculate that this could be a link to the problem of understanding higher-derivative corrections in DFT.}

Another reason to be optimistic about the existence of a non-linear extension is that in `exceptional field
theory' (the extension of DFT to U-duality groups) dual graviton components are already encoded
at the non-linear level \cite{Hohm:2013jma}, which is achieved by means of additional (compensator) fields.
The detailed formulation of these theories is somewhat different, however, in that they require
a split of coordinates and indices so that the mixed Young tableau nature of the dual graviton is no longer visible.
Therefore, the precise relation between the dual formulation presented here and that implicit in \cite{Hohm:2013jma}
remains to be established. Once this has been achieved and/or the full non-linear form
of the dual DFT has been constructed,  we would have a fully duality covariant formulation of
the low-energy dynamics of the type II strings in terms of all fields and their duals, both for the RR sector, for which this was established a while ago \cite{Hohm:2011zr,Jeon:2012hp}, and the NS sector.

The construction of such a theory would be very important for the description of various
types of (exotic) branes. Indeed, exotic branes are non-perturbative string states that are electrically charged with respect to mixed-symmetry potentials.
The branes that are charged under the $D$ potentials discussed in this paper have tensions that scale like $g_{s}^{-2}$ in string frame. While the NS5-brane is charged under the standard potential $D^{abcd}$, the KK monopole, the $Q$-brane and the $R$-brane are charged under the mixed-symmetry potentials $D^{abc}{}_d$, $D^{ab}{}_{cd}$ and $D^{a}{}_{bcd}$, respectively. The $Q$-brane solution \cite{deBoer:2012ma} is locally geometric, while the $R$-brane does not admit a geometric description. This is clearly in agreement with  our findings, namely that one can write down a duality relation in $x$ space at the linearized level for  $D^{ab}{}_{cd}$ but not for  $D^{a}{}_{bcd}$. Actually, one should also consider non-geometric objects that are charged under the potential $D_{abcd}$. Upon dimensional reduction, this would give rise to space-filling branes
with the same scaling of the tension, which have been classified (see the second ref.~in \cite{Bergshoeff:2010xc}). In general these branes do not have any solution in supergravity, but their existence is crucial for instance in orientifold models.

The 1/2-BPS branes with tension $g_s^{-2}$ satisfy specific `wrapping rules' \cite{Bergshoeff:2011mh}: the number of $p$-branes in $D$ dimensions is given by the number of $p+1$-branes in $D+1$ dimensions plus twice the number of $p$-branes in $D+1$ dimensions. This means that these branes `double' when they do not wrap the internal cycle. As far as the $(D-5)$-branes, the $(D-4)$-branes and the $(D-3)$-branes are concerned, this is expected from the fact that such branes are magnetically dual to the fundamental string, fundamental particles and fundamental instantons,  respectively. Therefore, for these branes the wrapping rules are simply the dual of the wrapping rules for fundamental strings, that see a doubled circle, and thus double when they wrap. The fact that all the potentials associated to these
branes enter the DFT duality relations discussed in this paper explains why also the $(D-2)$ and $(D-1)$-branes with tension proportional to $g_{s}^{-2}$ satisfy the same wrapping rules, although they are not dual to propagating fields in $x$ space.

The classification of 1/2-BPS branes in string theory was extended to branes with  tension scaling like $g_{s}^{-3}$ in the string frame in \cite{Bergshoeff:2011ee}. Such branes are charged with respect to  mixed-symmetry potentials that are magnetically dual to the $P$-fluxes (a prototype of a $P$-flux is the S-dual of the $Q$-flux). In \cite{Bergshoeff:2015cba} it was observed that  all such potentials can be collected in the  field $E_{MN, \dot{\alpha}}$ in the tensor-spinor representation of  $\text{SO}(10,10)$. It would be very interesting to write down a linearized DFT duality relation for such field, precisely as we did for the $D$ fields in this paper. Such field is magnetically dual to the  $p$-form potentials $\gamma_{a_1 ... a_p}$ (with $p$ even in IIB and odd in IIA), that are U-dual to the RR fields (for instance the IIB scalar $\gamma$ is the S-dual of the RR axion), and group together to form a spinor representation of $\text{SO}(10,10)$. The branes with tension $g_{s}^{-3}$ satisfy different wrapping rules with respect to the $g_{s}^{-2}$ branes, namely they `double' both if they wrap and if they do not wrap. The precise DFT duality relation between the potential  $E_{MN, \dot{\alpha}}$ and the potentials $\gamma$ would give an explanation for this wrapping rule, which is at the moment rather mysterious.

\vskip .7cm

\section*{Acknowledgments}


We are grateful to Nicolas Boulanger for helpful explanations on exotic dualizations.
V.A.P thanks Athanasios Chatzistavrakidis for enlightening  discussions. 
E.A.B. wishes to acknowledge the support and hospitality of the Center for Theoretical Physics at MIT where this work was started. E.A.B, V.A.P., F.R wish to acknowledge the support and hospitality of Mainz Institute for Theoretical Physics. F.R. thanks the Van Swinderen Institute, University of Groningen, for hospitality. 
O.H.~is supported by a DFG Heisenberg Fellowship
by the German Science Foundation (DFG).


\begin{thebibliography}{99}

\bibitem{Hull:2000zn}
  C.~M.~Hull,
  ``Strongly coupled gravity and duality,''
  Nucl.\ Phys.\ B {\bf 583}, 237 (2000)
  doi:10.1016/S0550-3213(00)00323-0
  [hep-th/0004195].

\bibitem{West:2001as}
  P.~C.~West,
  ``E(11) and M theory,''
  Class.\ Quant.\ Grav.\  {\bf 18}, 4443 (2001)
  doi:10.1088/0264-9381/18/21/305
  [hep-th/0104081].

\bibitem{Bekaert:2002uh}
  X.~Bekaert, N.~Boulanger and M.~Henneaux,
  ``Consistent deformations of dual formulations of linearized gravity: A No go result,''
  Phys.\ Rev.\ D {\bf 67}, 044010 (2003)
  doi:10.1103/PhysRevD.67.044010
  [hep-th/0210278].

\bibitem{Bekaert:2004dz}
  X.~Bekaert, N.~Boulanger and S.~Cnockaert,
  ``No self-interaction for two-column massless fields,''
  J.\ Math.\ Phys.\  {\bf 46}, 012303 (2005)
  doi:10.1063/1.1823032
  [hep-th/0407102].

\bibitem{Siegel:1993th}
  W.~Siegel,
  ``Superspace duality in low-energy superstrings,''
  Phys.\ Rev.\ D {\bf 48} (1993) 2826
  doi:10.1103/PhysRevD.48.2826
   [hep-th/9305073].

  W.~Siegel,
  ``Two vierbein formalism for string inspired axionic gravity,''
  Phys.\ Rev.\ D {\bf 47} (1993) 5453
  doi:10.1103/PhysRevD.47.5453
  [hep-th/9302036].

		
\bibitem{Hull:2009mi}
  C.~Hull and B.~Zwiebach,
  ``Double Field Theory,''
  JHEP {\bf 0909} (2009) 099
  doi:10.1088/1126-6708/2009/09/099
  [arXiv:0904.4664 [hep-th]].
	

  C.~Hull and B.~Zwiebach,
  ``The Gauge algebra of double field theory and Courant brackets,''
  JHEP {\bf 0909} (2009) 090
  doi:10.1088/1126-6708/2009/09/090
  [arXiv:0908.1792 [hep-th]].

	
  O.~Hohm, C.~Hull and B.~Zwiebach,
  ``Background independent action for double field theory,''
  JHEP {\bf 1007} (2010) 016
  doi:10.1007/JHEP07(2010)016
  [arXiv:1003.5027 [hep-th]].

	
\bibitem{Hohm:2010pp}
  O.~Hohm, C.~Hull and B.~Zwiebach,
  ``Generalized metric formulation of double field theory,''
  JHEP {\bf 1008} (2010) 008
  doi:10.1007/JHEP08(2010)008
  [arXiv:1006.4823 [hep-th]].

  O.~Hohm and S.~K.~Kwak,
  ``Frame-like Geometry of Double Field Theory,''
  J.\ Phys.\ A {\bf 44} (2011) 085404
  doi:10.1088/1751-8113/44/8/085404
  [arXiv:1011.4101 [hep-th]].

	
\bibitem{Aldazabal:2013sca}
  G.~Aldazabal, D.~Marques and C.~Nunez,
  ``Double Field Theory: A Pedagogical Review,''
  Class.\ Quant.\ Grav.\  {\bf 30} (2013) 163001
  doi:10.1088/0264-9381/30/16/163001
  [arXiv:1305.1907 [hep-th]].

  D.~S.~Berman and D.~C.~Thompson,
  ``Duality Symmetric String and M-Theory,''
  Phys.\ Rept.\  {\bf 566}, 1 (2014)
  doi:10.1016/j.physrep.2014.11.007
  [arXiv:1306.2643 [hep-th]].

  O.~Hohm, D.~L\"ust and B.~Zwiebach,
  ``The Spacetime of Double Field Theory: Review, Remarks, and Outlook,''
  Fortsch.\ Phys.\  {\bf 61}, 926 (2013)
  doi:10.1002/prop.201300024
  [arXiv:1309.2977 [hep-th]].
	
	
\bibitem{Eyras:1998hn}
  E.~Eyras, B.~Janssen and Y.~Lozano,
  ``Five-branes, K K monopoles and T duality,''
  Nucl.\ Phys.\ B {\bf 531} (1998) 275
  doi:10.1016/S0550-3213(98)00575-6
  [hep-th/9806169].


\bibitem{Bekaert:2002dt} 
  X.~Bekaert and N.~Boulanger,
  ``Tensor gauge fields in arbitrary representations of GL(D,R): Duality and Poincare lemma,''
  Commun.\ Math.\ Phys.\  {\bf 245}, 27 (2004)
  doi:10.1007/s00220-003-0995-1
  [hep-th/0208058].

\bibitem{deMedeiros:2002qpr}
  P.~de Medeiros and C.~Hull,
  ``Exotic tensor gauge theory and duality,''
  Commun.\ Math.\ Phys.\  {\bf 235}, 255 (2003)
  doi:10.1007/s00220-003-0810-z
  [hep-th/0208155].
  
\bibitem{Siegel:1986tn} 
  W.~Siegel and B.~Zwiebach,
  ``OSP(1,1/2) String Field Theory with the World Sheet Metric,''
  Nucl.\ Phys.\ B {\bf 288}, 332 (1987).
  doi:10.1016/0550-3213(87)90218-5 

\bibitem{Hull:2001iu}
  C.~M.~Hull,
  ``Duality in gravity and higher spin gauge fields,''
  JHEP {\bf 0109}, 027 (2001)
  doi:10.1088/1126-6708/2001/09/027
  [hep-th/0107149].
  
\bibitem{Deser:1980fy} 
  S.~Deser, P.~K.~Townsend and W.~Siegel,
  ``Higher Rank Representations of Lower Spin,''
  Nucl.\ Phys.\ B {\bf 184}, 333 (1981).
  doi:10.1016/0550-3213(81)90222-4 

	
\bibitem{Bergshoeff:2010xc}
  E.~A.~Bergshoeff and F.~Riccioni,
  ``D-Brane Wess-Zumino Terms and U-Duality,''
  JHEP {\bf 1011} (2010) 139
  doi:10.1007/JHEP11(2010)139
  [arXiv:1009.4657 [hep-th]].
	
		E.~A.~Bergshoeff and F.~Riccioni,
  ``String Solitons and T-duality,''
  JHEP {\bf 1105} (2011) 131
  doi:10.1007/JHEP05(2011)131
  [arXiv:1102.0934 [hep-th]].
	

  E.~A.~Bergshoeff, T.~Ortin and F.~Riccioni,
  ``Defect Branes,''
  Nucl.\ Phys.\ B {\bf 856} (2012) 210
  doi:10.1016/j.nuclphysb.2011.10.037
  [arXiv:1109.4484 [hep-th]].
	
	
\bibitem{Chatzistavrakidis:2013jqa}
  A.~Chatzistavrakidis, F.~F.~Gautason, G.~Moutsopoulos and M.~Zagermann,
  ``Effective actions of nongeometric five-branes,''
  Phys.\ Rev.\ D {\bf 89} (2014) 6,  066004
  doi:10.1103/PhysRevD.89.066004
  [arXiv:1309.2653 [hep-th]].

  A.~Chatzistavrakidis and F.~F.~Gautason,
  ``U-dual branes and mixed symmetry tensor fields,''
  Fortsch.\ Phys.\  {\bf 62} (2014) 743
  doi:10.1002/prop.201400023
  [arXiv:1404.7635 [hep-th]].

\bibitem{deBoer:2012ma}
  J.~de Boer and M.~Shigemori,
  ``Exotic Branes in String Theory,''
  Phys.\ Rept.\  {\bf 532} (2013) 65
  doi:10.1016/j.physrep.2013.07.003
  [arXiv:1209.6056 [hep-th]].

  J.~de Boer and M.~Shigemori,
  ``Exotic branes and non-geometric backgrounds,''
  Phys.\ Rev.\ Lett.\  {\bf 104} (2010) 251603
  doi:10.1103/PhysRevLett.104.251603
  [arXiv:1004.2521 [hep-th]].



\bibitem{Geissbuhler:2013uka}
  D.~Geissbuhler, D.~Marques, C.~Nunez and V.~Penas,
  ``Exploring Double Field Theory,''
  JHEP {\bf 1306} (2013) 101
  doi:10.1007/JHEP06(2013)101
  [arXiv:1304.1472 [hep-th]].


\bibitem{Geissbuhler:2011mx}
  D.~Geissbuhler,
  ``Double Field Theory and N=4 Gauged Supergravity,''
  JHEP {\bf 1111} (2011) 116
  doi:10.1007/JHEP11(2011)116
  [arXiv:1109.4280 [hep-th]].

  G.~Aldazabal, W.~Baron, D.~Marques and C.~Nunez,
  ``The effective action of Double Field Theory,''
  JHEP {\bf 1111} (2011) 052
   [JHEP {\bf 1111} (2011) 109]
  doi:10.1007/JHEP11(2011)052, 10.1007/JHEP11(2011)109
  [arXiv:1109.0290 [hep-th]].
  
  M.~Grana and D.~Marques,
  ``Gauged Double Field Theory,''
  JHEP {\bf 1204} (2012) 020
  doi:10.1007/JHEP04(2012)020
  


\bibitem{Bergshoeff:2015cba}
  E.~A.~Bergshoeff, V.~A.~Penas, F.~Riccioni and S.~Risoli,
  ``Non-geometric fluxes and mixed-symmetry potentials,''
  JHEP {\bf 1511} (2015) 020
  doi:10.1007/JHEP11(2015)020
  [arXiv:1508.00780 [hep-th]].

\bibitem{Boulanger:2012df}
  N.~Boulanger, P.~P.~Cook and D.~Ponomarev,
  ``Off-Shell Hodge Dualities in Linearised Gravity and E11,''
  JHEP {\bf 1209}, 089 (2012)
  doi:10.1007/JHEP09(2012)089
  [arXiv:1205.2277 [hep-th]].

  N.~Boulanger and D.~Ponomarev,
  ``Frame-like off-shell dualisation for mixed-symmetry gauge fields,''
  J.\ Phys.\ A {\bf 46} (2013) 214014
  doi:10.1088/1751-8113/46/21/214014
  [arXiv:1206.2052 [hep-th]].

  N.~Boulanger, P.~Sundell and P.~West,
  ``Gauge fields and infinite chains of dualities,''
  arXiv:1502.07909 [hep-th].


\bibitem{Hohm:2011dz} 
  O.~Hohm,
  ``On factorizations in perturbative quantum gravity,''
  JHEP {\bf 1104}, 103 (2011)
  doi:10.1007/JHEP04(2011)103
  [arXiv:1103.0032 [hep-th]].
  
  O.~Hohm and D.~Marques,
  ``Perturbative Double Field Theory on General Backgrounds,''
  Phys.\ Rev.\ D {\bf 93}, no. 2, 025032 (2016)
  doi:10.1103/PhysRevD.93.025032
  [arXiv:1512.02658 [hep-th]].  

\bibitem{Hohm:2013jma}
  O.~Hohm and H.~Samtleben,
  ``U-duality covariant gravity,''
  JHEP {\bf 1309}, 080 (2013)
  doi:10.1007/JHEP09(2013)080
  [arXiv:1307.0509 [hep-th]].

  O.~Hohm and H.~Samtleben,
  ``Exceptional Form of D=11 Supergravity,''
  Phys.\ Rev.\ Lett.\  {\bf 111}, 231601 (2013)
  doi:10.1103/PhysRevLett.111.231601
  [arXiv:1308.1673 [hep-th]].

 O.~Hohm and H.~Samtleben,
  ``Exceptional Field Theory I: $E_{6(6)}$ covariant Form of M-Theory and Type IIB,''
  Phys.\ Rev.\ D {\bf 89}, no. 6, 066016 (2014)
  doi:10.1103/PhysRevD.89.066016
  [arXiv:1312.0614 [hep-th]].

  O.~Hohm and H.~Samtleben,
  ``Exceptional field theory. II. E$_{7(7)}$,''
  Phys.\ Rev.\ D {\bf 89}, 066017 (2014)
  doi:10.1103/PhysRevD.89.066017
  [arXiv:1312.4542 [hep-th]].

  O.~Hohm and H.~Samtleben,
  ``Exceptional field theory. III. E$_{8(8)}$,''
  Phys.\ Rev.\ D {\bf 90}, 066002 (2014)
  doi:10.1103/PhysRevD.90.066002
  [arXiv:1406.3348 [hep-th]].

\bibitem{Godazgar:2014nqa}
  H.~Godazgar, M.~Godazgar, O.~Hohm, H.~Nicolai and H.~Samtleben,
  ``Supersymmetric E$_{7(7)}$ Exceptional Field Theory,''
  JHEP {\bf 1409}, 044 (2014)
  doi:10.1007/JHEP09(2014)044
  [arXiv:1406.3235 [hep-th]].

\bibitem{Hohm:2014qga}
  O.~Hohm and H.~Samtleben,
  ``Consistent Kaluza-Klein Truncations via Exceptional Field Theory,''
  JHEP {\bf 1501}, 131 (2015)
  doi:10.1007/JHEP01(2015)131
  [arXiv:1410.8145 [hep-th]].


\bibitem{Boulanger:2003vs} 
  N.~Boulanger, S.~Cnockaert and M.~Henneaux,
  ``A note on spin s duality,''
  JHEP {\bf 0306}, 060 (2003)
  doi:10.1088/1126-6708/2003/06/060
  [hep-th/0306023].


\bibitem{Andriot:2012an}

  D.~Andriot, M.~Larfors, D.~Lust and P.~Patalong,
  ``A ten-dimensional action for non-geometric fluxes,''
  JHEP {\bf 1109} (2011) 134
  doi:10.1007/JHEP09(2011)134
  [arXiv:1106.4015 [hep-th]].


  D.~Andriot, O.~Hohm, M.~Larfors, D.~Lust and P.~Patalong,
  ``A geometric action for non-geometric fluxes,''
  Phys.\ Rev.\ Lett.\  {\bf 108} (2012) 261602
  doi:10.1103/PhysRevLett.108.261602
  [arXiv:1202.3060 [hep-th]].


  D.~Andriot, O.~Hohm, M.~Larfors, D.~Lust and P.~Patalong,
  ``Non-Geometric Fluxes in Supergravity and Double Field Theory,''
  Fortsch.\ Phys.\  {\bf 60} (2012) 1150
  doi:10.1002/prop.201200085
  [arXiv:1204.1979 [hep-th]].
	
	
  D.~Andriot and A.~Betz,
  ``$\beta$-supergravity: a ten-dimensional theory with non-geometric fluxes, and its geometric framework,''
  JHEP {\bf 1312} (2013) 083
  doi:10.1007/JHEP12(2013)083
  [arXiv:1306.4381 [hep-th]].
  	





\bibitem{Hohm:2012mf}

  O.~Hohm and B.~Zwiebach,
  ``On the Riemann Tensor in Double Field Theory,''
  JHEP {\bf 1205} (2012) 126
  doi:10.1007/JHEP05(2012)126
  [arXiv:1112.5296 [hep-th]].

	
	O.~Hohm and B.~Zwiebach,
  ``Towards an invariant geometry of double field theory,''
  J.\ Math.\ Phys.\  {\bf 54} (2013) 032303
  doi:10.1063/1.4795513
  [arXiv:1212.1736 [hep-th]].


\bibitem{Hohm:2013jaa}

  O.~Hohm, W.~Siegel and B.~Zwiebach,
  ``Doubled $\alpha'$-geometry,''
  JHEP {\bf 1402}, 065 (2014)
  doi:10.1007/JHEP02(2014)065
  [arXiv:1306.2970 [hep-th]].

  O.~Hohm and B.~Zwiebach,
  ``Double field theory at order $\alpha'$,''
  JHEP {\bf 1411}, 075 (2014)
  doi:10.1007/JHEP11(2014)075
  [arXiv:1407.3803 [hep-th]].
	
	
\bibitem{Hohm:2011zr}
  O.~Hohm, S.~K.~Kwak and B.~Zwiebach,
  ``Unification of Type II Strings and T-duality,''
  Phys.\ Rev.\ Lett.\  {\bf 107}, 171603 (2011)
  doi:10.1103/PhysRevLett.107.171603
  [arXiv:1106.5452 [hep-th]], ``Double Field Theory of Type II Strings,''
  JHEP {\bf 1109}, 013 (2011)
  doi:10.1007/JHEP09(2011)013
  [arXiv:1107.0008 [hep-th]].
	
\bibitem{Jeon:2012hp} 
  I.~Jeon, K.~Lee, J.~H.~Park and Y.~Suh,
  ``Stringy Unification of Type IIA and IIB Supergravities under N=2 D=10 Supersymmetric Double Field Theory,''
  Phys.\ Lett.\ B {\bf 723}, 245 (2013)
  doi:10.1016/j.physletb.2013.05.016
  [arXiv:1210.5078 [hep-th]].	
	
\bibitem{Bergshoeff:2011mh} 
  E.~A.~Bergshoeff and F.~Riccioni,
  ``Dual doubled geometry,''
  Phys.\ Lett.\ B {\bf 702}, 281 (2011)
  doi:10.1016/j.physletb.2011.07.009
  [arXiv:1106.0212 [hep-th]].
  
\bibitem{Bergshoeff:2011ee} 
  E.~A.~Bergshoeff and F.~Riccioni,
  ``Branes and wrapping rules,''
  Phys.\ Lett.\ B {\bf 704}, 367 (2011)
  doi:10.1016/j.physletb.2011.09.043
  [arXiv:1108.5067 [hep-th]].
	

\end{thebibliography}
\end{document}